\documentclass{jfm}

\usepackage{graphicx}
\usepackage{newtxtext}
\usepackage{newtxmath}
\usepackage{natbib}
\usepackage{hyperref}
\usepackage{caption}
\usepackage{subcaption}
\usepackage{ dsfont }
\hypersetup{
    colorlinks = true,
    urlcolor   = blue,
    citecolor  = black,
}

\newcommand\cpc[1]{{#1}}
\newcommand\np[1]{{#1}}
\newcommand\rev[1]{{#1}}

\newcommand\Ast[1]{{{#1}^{\ast}}}

\newcommand{\RomanNumeralCaps}[1]
\nolinenumbers

\title{Turbulent disruption of density staircases in stratified shear flows}

\shorttitle{Turbulent disruption of density staircases}

\author{Nicolaos Petropoulos\aff{1}
  \corresp{\email{np546@cam.ac.uk}}, Ali Mashayek\aff{2}
 \and Colm-cille P. Caulfield\aff{3,1}}

\affiliation{\aff{1}Department of Applied Mathematics and Theoretical Physics, University of Cambridge, Cambridge CB3 0WA, UK
\aff{2}Department of Civil and Environmental Engineering, Imperial College London, London SW7 2BX, UK
\aff{3}Institute for Energy and Environmental Flows, University of Cambridge, Cambridge CB3 0EZ, UK}

\begin{document}
\maketitle

\begin{abstract}
Formation of step-like `density staircase' distributions induced by stratification and turbulence has been  widely studied and can be explained by the `instability' of a sufficiently strongly stably stratified turbulent flow due to the decrease of the turbulent density flux with increasing stratification via the `Phillips mechanism'\citep{phillips1972turbulence}. However, such density staircases are not often observed in ocean interiors, except in regions where double diffusion processes are important, leading to thermohaline staircases. Using reduced order models for the evolution of velocity and density gradients, we analyse staircase formation in stratified and sheared turbulent flows. Under the assumption of inertial scaling $\epsilon \sim U^3/L$ for the  kinetic energy dissipation rate $\epsilon$, where $U$ and $L$ are characteristic velocity and length scales, we determine ranges of bulk Richardson numbers $\text{Ri}_{b}$ and turbulent Prandtl numbers $\text{Pr}_{T}$ for which staircases can potentially form and show that the Phillips mechanism only survives in the limit of sufficiently small turbulent Prandtl numbers. For relevant oceanic parameters, a \rev{range of turbulent Prandtl numbers} above which the system is not prone to staircases is found to be $\text{Pr}_T \simeq 0.5 - 0.8$. Since several studies indicate that the turbulent Prandtl number in stably stratified turbulence and in ocean interiors is usually above this threshold, this result supports the empirical observation that staircases are not favoured in ocean interiors in the presence of \rev{relatively homogeneous and sustained turbulence}. We also show that our analysis is robust to other scalings for $\epsilon$ (such as the more strongly stratified scaling $\epsilon \sim U^{2}N_{c}$, where $N_{c}$ is a characteristic value of the buoyancy frequency), supporting our results in both shear-dominated and buoyancy-dominated turbulent regimes as well as in weakly and strongly stratified regimes. 
\end{abstract}



\section{Introduction}
\label{section:Introduction}

Spontaneous formation of step-like `density staircases'  distributions \--- made up of a series of \cpc{relatively deep and well-mixed `layers' separated by relatively thin `interfaces' of enhanced density gradient} 
\--- induced by stratification and turbulence has been postulated and studied by many authors \citep{phillips1972turbulence, posmentier1977generation} and has been observed in a number of contexts. Experimentally, density staircases form when dragging a rod or a grid through a stable salt gradient \citep{Linden_1980, thorpe_1982, ruddick1989formation, park1994turbulent} or in stratified turbulent Taylor-Couette flows \citep{oglethorpe2013spontaneous}. In the oceans, they have been detected in regions where double-diffusion is important (in polar regions for \cpc{example}) and leads to the development of thermohaline staircases \citep{timmermans2008ice, radko_2016}. In the Arctic, their presence is crucial, \cpc{not least because} they act as a barrier to mixing, protecting the Arctic ice from the heat input inflowing from the Atlantic ocean~\citep{rippeth2022turbulent}. \cpc{In} astrophysical stratified flows, density \cpc{staircases could} potentially form in regions with \cpc{sufficiently} large molecular Prandtl number ($\mathcal{O}(10^{-3})$ or larger; in white dwarf interiors for \cpc{example}) thanks to fingering convection~\citep{garaud2015excitation}. The formation of such structures is due to the interaction between small scale turbulence and larger scale stratification. \cpc{Such} turbulence is inherently anisotropic as stratification tends to inhibit vertical motions, and inhomogeneous \cpc{due to the inevitable} presence of sharp density interfaces. 

\cpc{Although stratified turbulence is thus inevitably} difficult to analyse, insight has been gained using flux-gradient parameterisations. Using such models, \citet{phillips1972turbulence} and \citet{posmentier1977generation} reduced the dynamics of the staircase formation problem (with \cpc{a single} stratification agent) to the following nonlinear diffusion equation for the horizontally averaged buoyancy $\Bar{b}$: 
\begin{equation}
    \partial_{t}\Bar{b} = \partial_{z}[F(\partial_{z}\Bar{b})],  
\end{equation}
where, importantly, the turbulent buoyancy flux $F$ is a non-monotonic function of the buoyancy gradient \citep{Linden_1979}. Using this formulation, \rev{staircase formation} can be explained by an `instability'  of a sufficiently strongly stably stratified turbulent flow due to the decrease of the turbulent buoyancy flux with increasing stratification,
\cpc{through what is now commonly referred to as the `Phillips mechanism'}. Flux-gradient parameterisations have, however, some drawbacks. Firstly, they are antidiffusive when the flux is a decreasing function of the gradient, leading to mathematically ill-posed problems, \cpc{and it is this ill-posedness which manifests itself as the `instability' of the Phillips mechanism.}  Secondly, they are not valid at all scales and tend to break down when the size of the phenomenon of interest (the layers in our case) is of \cpc{the} order \cpc{of} magnitude or smaller \cpc{of} the turbulent microstructures that such models try to parameterise \citep{radko_2014}. These issues can both be resolved using the recently developed multi-scale analysis introduced by~\citet{radko_2019} in the context of thermohaline staircase formation. Carefully introducing the interplay between different scales into the flux-gradient parameterisations, this method corrects the models at small scales and generates mathematically well-posed systems. Other \cpc{regularisation} techniques have also been proposed. \citet{barenblatt1993mathematical} used a time-delayed flux-gradient model to construct a mathematically well-posed model of mixing in stratified turbulent flows whereas~\citet{balmforth1998dynamics} considered the evolution of both buoyancy gradients and turbulent kinetic energy to analyse \cpc{staircase} dynamics in stratified turbulent flows. 

The above reduced order model predicts \rev{staircase} formation for sufficiently strongly stratified flows. \cpc{Furthermore,} \citet{billant2001self} identified a strongly stratified regime (in the sense that the horizontal Froude number $\text{Fr}_{h} = U/L_{h}N_{c}$ is small, where $U$ is a characteristic  horizontal velocity \cpc{scale}, $L_{h}$ is a typical horizontal length scale, and $N_{c}$ is a characteristic value of \cpc{the buoyancy frequency}) for which the (full) equations of motion are self-similar with respect to $zN_{c}/U$, suggesting a layered structure with characteristic \cpc{vertical} length scale $U/N_{c}$. These vertical \rev{staircases} offer a route for turbulence to grow and be sustained in strongly stratified flows and hence mix strong density gradients. Indeed, whereas sufficiently weakly stratified flows are prone to shear instabilities that can overturn density gradients, strongly stratified flows prevent \cpc{such} instabilities from growing. However, they are prone to \rev{staircase formation} that reduces (locally) the stratification inside the layers, creating a favourable environment for shear instabilities to develop~\citep{cope_garaud_caulfield_2020}. The subsequent turbulence \cpc{is inevitably} spatially and temporally intermittent and \cpc{is} characterised by scouring dynamics near the \cpc{relatively sharp} density interfaces rather than overturns, emphasizing the qualitatively different mixing expected in relatively weakly or strongly stratified flows~\citep{woods2010non, Annurev_Colm_Caulfield}. 

\begin{figure}
    \centerline{
    \includegraphics[width=\linewidth, trim={0 0 0 12cm}]{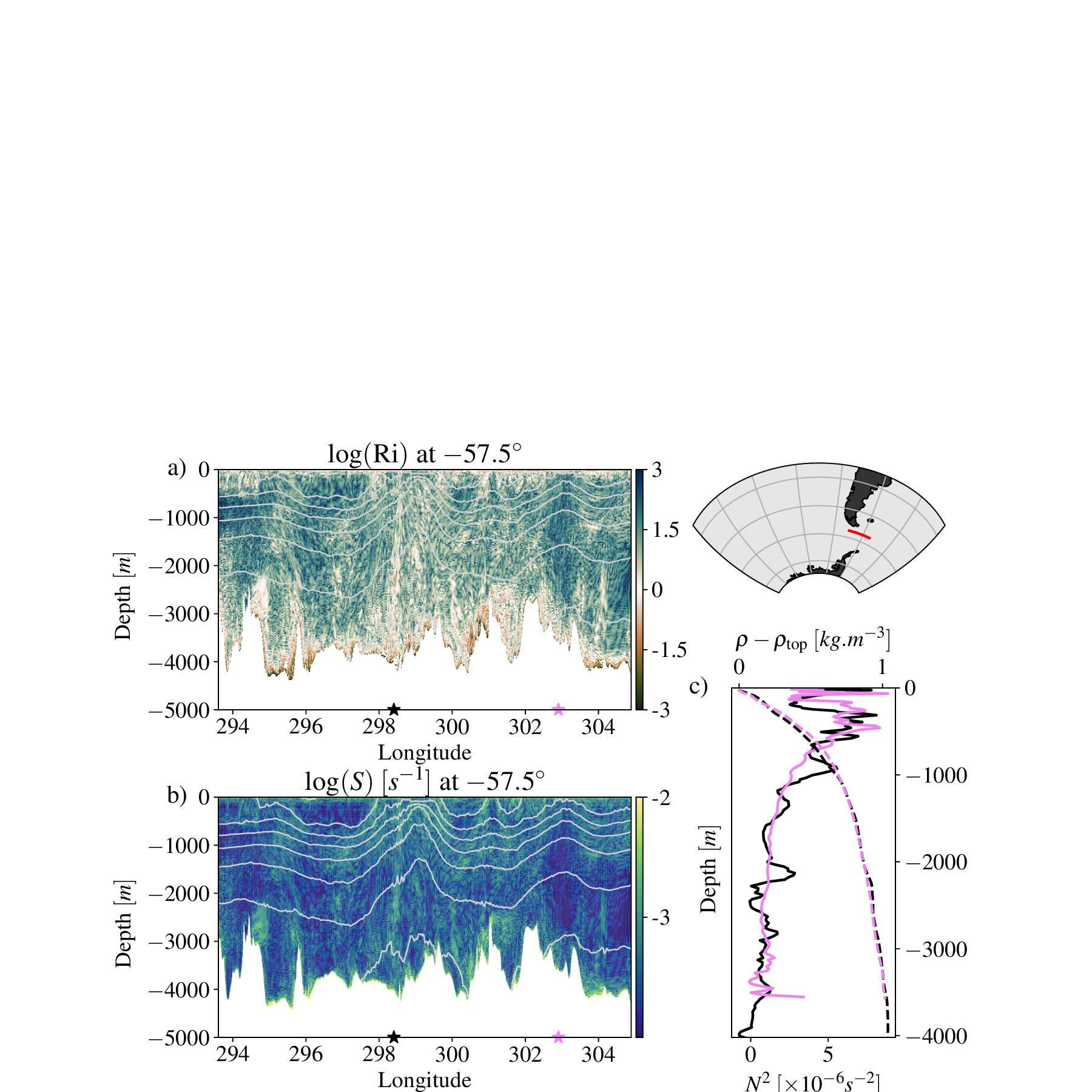}}
    \caption{(Panel a) The Richardson number on a slice in the Drake Passage of the Southern Ocean, corresponding to the red line in the inset on the top right corner. Buoyancy levels are overlaid in form of grey lines. (Panel b) Same as panel $a$ but for velocity shear. (Panel c) Density (dashed lines) and stratification (solid lines) profiles corresponding to the longitudes marked by stars in panels $a$ and $b$. Reproduced from~\citep{mashayek2022marginal}. } 
    \label{fig:intro}
\end{figure}

\rev{Oceanic flows are indeed often strongly stratified in the sense that an appropriate gradient Richardson number $\text{Ri}$, (defined here as the square of the ratio of the background buoyancy frequency and background vertical shear) is large enough 
so that
the Richardson number falls on the right flank of the turbulent buoyancy flux curve~\citep{Linden_1979}. As an example, figure~\ref{fig:intro}, reproduced from~\citep{mashayek2022marginal}, shows emergence of turbulence in the otherwise quiescent ocean interior when shear (from internal waves, mesoscale instabilities or boundary processes for instance; see figure~\ref{fig:intro}, panel b)) increases sufficiently for the Richardson number to drop below critical values. In the close vicinity of top and boundary turbulence is less intermittent. Such turbulent patches in the interior, sufficiently far from the boundaries, typically correspond to buoyancy gradients on the decreasing flank of the aforementioned turbulent buoyancy flux curve. Layering should on the face of it play an important role in formation and erosion of density gradients. However, panel c) in figure~\ref{fig:intro} shows that turbulent patches in the interior do not leave the density structure layered. This behavior seems generic in many parts of the ocean interior, of course excluding regions where thermohaline diffusive processes (e.g. double diffusion) can play a prominent role such as in the Arctic ocean or the Mediterranean Sea~\citep{timmermans2008ice, radko_2016}. Crucially, the shear and its spatiotemporal variability are key to turbulent mixing, yet absent from the theoretical framework that forms the basis for the Phillips mechanism. }

Motivated by these observations, in this work we analyse staircase formation (or lack thereof) in density stratified turbulence in presence of velocity shear (e.g. the interior turbulent patches mentioned above), and assess in which regime(s) it is possible for the Phillips mechanism \rev{\--- defined here as the instability with respect to small perturbations of linear buoyancy profiles in a turbulent flow far from boundaries~\citep{phillips1972turbulence} \---} to survive. Using reduced order models for the evolution of velocity and density gradients based on flux-gradient parameterisations of the turbulent fluxes (corrected using a simpler version of~\citet{radko_2019} multi-scale analysis) and under various scalings for the rate of dissipation of the kinetic energy $\epsilon$ (specifically $\epsilon \sim U^{3}/L$ and $\epsilon \sim U^{2}N_{c}$ where $L$ is a characteristic length-scale of our problem), we determine ranges of bulk Richardson numbers $\text{Ri}_{b}$ and turbulent Prandtl numbers $\text{Pr}_{T}$ \cpc{(defined more precisely below, effectively quantifying the relative strength of velocity shear to the buoyancy frequency and the ratio of turbulent diffusivity of momentum to turbulent diffusivity of buoyancy respectively)} for which staircases can potentially form. 

We demonstrate that the Phillips mechanism for \rev{staircase} formation in strongly stratified flows remains viable in the presence of shear only in the limit $\text{Pr}_{T} \ll 1$ but breaks down otherwise. \cpc{Specifically}, for \cpc{sufficiently large} $\text{Ri}_{b} \gtrsim 1$ there exists a limiting value of the turbulent Prandtl number $\text{Pr}_{T}$ above which \rev{staircase formation} \cpc{via this mechanism ceases to be possible}. For relevant oceanic parameters, this value is found around $0.5 - 0.8$. \np{Even though it is still challenging to measure the turbulent Prandtl number in the oceans,} several studies of direct numerical simulation of stably stratified turbulence indicate that $\text{Pr}_T$ 
is \cpc{typically non-trivially} above this threshold \citep{shih_koseff_ivey_ferziger_2005, venayagamoorthy2010turbulent} and therefore \cpc{our} result supports \cpc{and explains} the empirical observation that staircases are not favoured in ocean interiors in the presence of \rev{relatively homogeneous and sustained turbulence} driven by velocity shears. 


\cpc{To demonstrate this key result, the rest of the paper is organised as follows. In} section~\ref{section:formulation} we introduce the theoretical model used throughout the paper to analyse staircase formation in both density (stably) stratified and sheared turbulent flows, \cpc{through extending} the work of~\citet{phillips1972turbulence} and~\citet{posmentier1977generation} to take into account the evolution of shear, and \cpc{define relevant dimensionless parameters.} In section~\ref{section:linear_stability_analysis} we describe the regions in the parameter space that are prone to staircase instabilities through a linear stability analysis of the governing equations. In section~\ref{section:instability_properties} we present some properties of the \cpc{various instabilities, while in} section~\ref{section:nonlinear_dynamics} we compare the nonlinear dynamics leading to \rev{staircase} formation and the  stability \cpc{analyses}. \cpc{Finally, we draw brief conclusions} in section~\ref{section:discussion}. 

\section{Formulation}
\label{section:formulation}

Except when stated otherwise, the following notations will be used throughout this work: 
\begin{itemize}
\setlength\itemsep{0.1em}
    \item a star $\Ast{\boldsymbol{\cdot}}$ denotes a dimensional variable. The star is dropped for dimensionless quantities;  
    \item an overbar $\Bar{\boldsymbol{\cdot}}$ denotes an horizontally averaged quantity; 
    \item a tilde $\Tilde{\boldsymbol{\cdot}}$ denotes a deviation from a background quantity; 
    \item a prime $\boldsymbol{\cdot}'$ denotes a derivative with respect to argument (always in fact being $\text{Ri}_{b}$). 
\end{itemize}

\subsection{Dimensional form}

\cpc{The} Navier-Stokes equations in the Boussinesq approximation (with a background density $\rho_{0}^{\ast}$) are: 
\begin{equation}
    \begin{cases}
    \partial_{\Ast{t}}\Ast{u} + \boldsymbol{\Ast{u}}\bcdot\bnabla^{\boldsymbol{\ast}} \Ast{u} = \nu^\ast\Ast{\nabla}^{2}\Ast{u} - \frac{1}{\rho_{0}}\partial_{\Ast{x}}\Ast{p}, \quad
    \partial_{\Ast{t}}\Ast{v} + \boldsymbol{\Ast{u}}\bcdot\bnabla^{\boldsymbol{\ast}} \Ast{v} = \nu^\ast\Ast{\nabla}^{2}\Ast{v} - \frac{1}{\rho_{0}^{\ast}}\partial_{\Ast{y}}\Ast{p}, \\
    \partial_{\Ast{t}}\Ast{w} + \boldsymbol{\Ast{u}}\bcdot\bnabla^{\boldsymbol{\ast}} \Ast{w} = \nu^\ast\Ast{\nabla}^{2}\Ast{w} - \frac{1}{\rho_{0}^{\ast}}\partial_{\Ast{z}}\Ast{p} + \Ast{b}, \quad
    \partial_{\Ast{t}}\Ast{b} + \boldsymbol{\Ast{u}}\bcdot\bnabla^{\boldsymbol{\ast}} \Ast{b} = \kappa^\ast\Ast{\nabla}^{2}\Ast{b}, \\
    \bnabla^{\boldsymbol{\ast}} \bcdot \boldsymbol{\Ast{u}} = 0, 
    \end{cases}
\end{equation}
where $\boldsymbol{\Ast{u}} = (\Ast{u},\Ast{v},\Ast{w})$ is the velocity field, $\Ast{b} := -\frac{\Ast{g}}{\rho_{0}^{\ast}}\Ast{\rho}$ is buoyancy (where $\Ast{\rho}$ is density and $\Ast{g}$ is the gravitational acceleration), $\Ast{p}$ is pressure and $\kappa^\ast$ and $\nu^\ast$ are the (molecular) diffusivity and viscosity of the fluid. The differential operators are taken with respect to dimensional quantities. Averaging  in the horizontal and assuming that $\boldsymbol{\Ast{u}} = (\overline{\Ast{u}}(z,t),0,0) + \boldsymbol{\Tilde{\Ast{u}}}$ and $\Ast{b} = \overline{\Ast{b}}(z,t) + \Tilde{\Ast{b}}$ where $\overline{\Ast{u}}$ and $\overline{\Ast{b}}$ are the horizontally averaged velocity and buoyancy profiles respectively, we obtain: 
\begin{equation}
    \begin{cases}
        \partial_{\Ast{t}} \overline{\Ast{b}} = \kappa^\ast\partial_{\Ast{z}}^{2}\overline{\Ast{b}} - \partial_{\Ast{z}}F_{b}^{\ast}, \; F_{b}^{\ast} = \overline{\Tilde{\Ast{b}}\Tilde{\Ast{w}}}, \\
        \partial_{\Ast{t}} \overline{\Ast{u}} = \nu^\ast\partial_{\Ast{z}}^{2}\overline{\Ast{u}} - \partial_{\Ast{z}}F_{u}^{\ast}, \; F_{u}^{\ast} = \overline{\Tilde{\Ast{u}}\Tilde{\Ast{w}}},   
    \end{cases}
    \label{eq:coupled_system_molecular_diffusivities}
\end{equation}
where $F_{b}^{\ast}$ and $F_{u}^{\ast}$ are respectively the vertical buoyancy and momentum turbulent fluxes, with overbars denoting horizontal averages. \cpc{Using gradient-flux models to parameterise these fluxes in terms of the mean  buoyancy and velocity gradients we obtain implicit definitions for the turbulent diffusivities of buoyancy $\kappa_T^{\ast}$ and momentum $\nu_T^{\ast}$:} 
\begin{equation}
    F_{b}^{\ast} = - \kappa^\ast_{T}\partial_{\Ast{z}}\overline{\Ast{b}}, \; F_{u}^{\ast} = - \nu^\ast_{T}\partial_{\Ast{z}}\overline{\Ast{u}}.
\end{equation}

Our goal is to understand how an ambient shear influences the formation of \cpc{density (or equivalently buoyancy) staircases. Therefore, we choose to model these fluxes only in terms of the gradient Richardson number $\text{Ri}_{g}$, defined in terms of the background shear $\Ast{S}$ and buoyancy frequency $\Ast{N}$:
 \begin{equation}
  \text{Ri}_{g} := \frac{\Ast{N}^{2}}{\Ast{S}^{2}}; 
  \ \Ast{S} := \partial_{\Ast{z}}\overline{\Ast{u}}, \ 
  \Ast{N}^{2} := \partial_{\Ast{z}}\overline{\Ast{b}}. \label{eq:ridef}
 \end{equation}
 }
 
\cpc{It is important to remember that  common parameterisations of the turbulent diffusivities  rely also on the buoyancy Reynolds number $\text{Re}_{b} := \Ast{\epsilon} / \Ast{\nu} \Ast{N}^{2}$ where $\Ast{\epsilon}$ is the dissipation rate of  turbulent kinetic energy \citep{shih_koseff_ivey_ferziger_2005, bouffard2013diapycnal, mashayek2017efficiency}. }
\cpc{We can however express the buoyancy flux as:
\begin{equation}
    F_{b}^{\ast} = - \Gamma \Ast{\epsilon} , \label{eq:gammadef}
\end{equation}
}where $\Gamma$ is the turbulent flux coefficient \citep{osborn1980estimates} and reduce the modeling of the turbulent diffusivities $\kappa^\ast_{T}$ and $\nu^\ast_{T}$ to the modeling of this coefficient. Parameterisations of $\Gamma$ in terms of $\text{Ri}_{g}$ have been presented in \citep{TheRelationshipbetweenFluxCoefficientandEntrainmentRatioinDensityCurrents} for instance. At $\text{Ri}_{g} = 0$, there is no buoyancy to mix and therefore it seems reasonable to assume $\Gamma(\text{Ri}_{g}=0) = 0$. As $\text{Ri}_{g}$ increases, there is more and more scalar to mix and $\Gamma$ should therefore increase. However, as stratification becomes more significant, it \cpc{is reasonable to suppose that it will suppress vertical motion because of restoring buoyancy forces, possibly leading to less efficient mixing. Whether $\Gamma$ decreases towards $0$ or saturates for $\text{Ri}_{g}$ large enough is still an open question \citep{Annurev_Colm_Caulfield}. However, the analysis presented in the \cpc{following sections depends most strongly on}  the monotonicity of the flux coefficient in terms of the Richardson number, and not the specific functional form of  $\Gamma(\text{Ri}_{g})$} \np{and hence the two cases can be studied, as we will see later. } 

Written in terms of the flux coefficient $\Gamma$, the mean buoyancy and velocity equations~\eqref{eq:coupled_system_molecular_diffusivities} are:
\begin{equation}
    \begin{cases}
        \partial_{\Ast{t}} \Ast{N}^2 = \kappa^\ast\partial_{\Ast{z}}^{2}\Ast{N}^2 + \partial_{\Ast{z}}^{2}[\Gamma\Ast{\epsilon}], \\
        \partial_{\Ast{t}} \Ast{S} = \nu^\ast\partial_{\Ast{z}}^{2}\Ast{S} + \text{Pr}_{T}\partial_{\Ast{z}}^{2}\left[\frac{\Gamma\Ast{\epsilon}}{\Ast{S}\text{Ri}_{g}}\right], \  \text{Pr}_{T} := \frac{\nu^\ast_{T}}{ \kappa^\ast_{T}}.   
    \end{cases}
    \label{eq:coupled_system_flux_coefficient}
\end{equation}
\rev{For the sake of simplicity and because our goal is to understand how this coupling parameter affects the formation of staircases, we consider the turbulent Prandtl number $\text{Pr}_T$ as a free constant parameter that does not depend on the stratification nor on the shear. } The above equations are coupled through the dependence of the flux coefficient $\Gamma$ on the Richardson number $\text{Ri}_{g}$. 
Moreover, since the system is invariant under the mapping $\Ast{S} \rightarrow -\Ast{S}$, we will assume without loss of generality that $\Ast{S} \geq 0$. Since we are considering statically stable buoyancy profiles, we also have $\Ast{N}^{2} \geq 0$. 

\begin{figure}
    \centerline{
    \includegraphics[width=\linewidth, trim={0 0 0 1.5cm}]{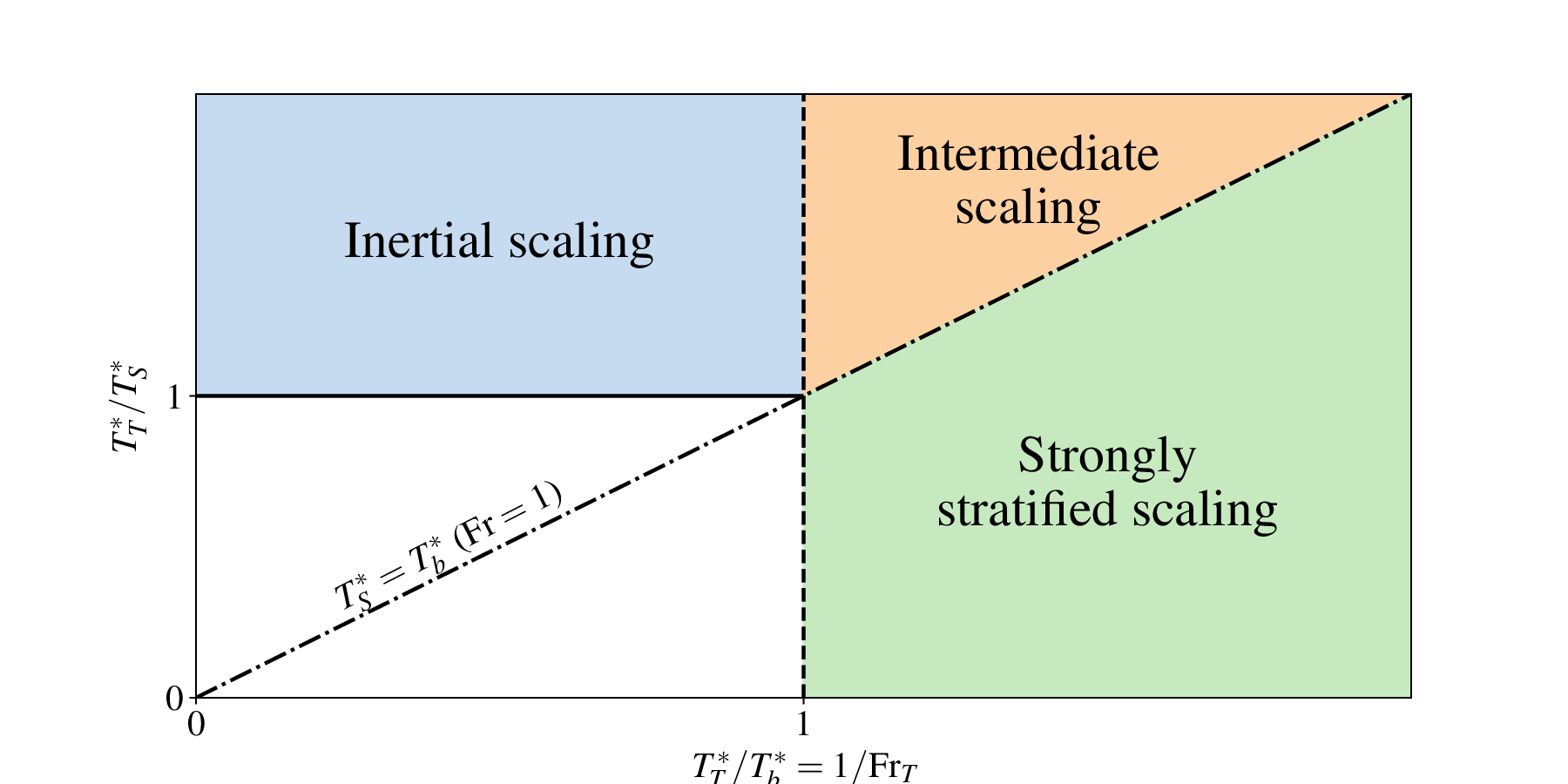}}
    \caption{Scaling used depending on the relative size of $T_{b}^{\ast}$ (buoyancy time-scale), $T_{S}^{\ast}$ (shear time-scale) and $T_{T}^{\ast}$ (turbulent time-scale). The solid horizontal line corresponds to $T_{T}^{\ast}/T_{S}^{\ast} = 1$. The dashed vertical line corresponds to $T_{T}^{\ast}/T_{b}^{\ast} = 1$. The dash-dotted line correspond to $T_{b}^{\ast}/T_{S}^{\ast} = 1$. }
    \label{fig:scalings}
\end{figure}

\subsection{Dimensionless system}
\label{section:non_dimensional_system}

In order to scale the system~\eqref{eq:coupled_system_flux_coefficient}, we need to make some assumptions regarding the relevant time-scale of our problem as well as on the dissipation rate of turbulent kinetic energy $\Ast{\epsilon}$. Using data from various sources, \citet{mater2014unifying} show that stably stratified shear-flow turbulence could be interpreted in terms of three time-scales: the buoyancy time-scale $T^\ast_{b} := 1/\Ast{N}$, the shear time-scale $T^\ast_{S} := 1/\Ast{S}$ and turbulence time-scale $T^\ast_{T} := \Ast{{\cal K}} / \Ast{\epsilon}$ where $\Ast{{\cal K}}$ is the turbulent kinetic energy density. In the following, we propose three different scalings that depend on the relative size of these time-scales. These scalings are summarized in figure~\ref{fig:scalings}.

\subsubsection{Inertial scaling}
\label{section:inertial_scaling}

\cpc{We first propose to scale the} system~\eqref{eq:coupled_system_flux_coefficient} \cpc{under} the assumption that the dissipation rate of turbulent kinetic energy $\Ast{\epsilon}$ \cpc{scales `inertially' like} $\Ast{U}^3/L^\ast$ where $U^\ast$ is a characteristic velocity \cpc{scale} and $L^\ast$ is a \cpc{characteristic length scale} of our problem. This scaling has been justified in many experimental and observational settings \citep{OntheNatureofTurbulenceinaStratifiedFluidPartITheEnergeticsofMixing, Ivey_1998, Kay_&_Jay_2003, shih_koseff_ivey_ferziger_2005}. 
\rev{It is relevant, for instance, in sufficiently weakly stratified or shear-dominated turbulent flows where the turbulent Froude number $\text{Fr}_{T} := \Ast{\epsilon} / \Ast{N} \Ast{{\cal K}} (= T_{b}^{\ast}/T_{T}^{\ast})$ as well as $T_{T}^{\ast}/T_{S}^{\ast}$ are sufficiently large (implying $\text{Fr} := \Ast{S}/\Ast{N} (= T_{b}^{\ast}/T_{S}^{\ast})$ sufficiently large) \citep{mater2014unifying}. Then, the relevant time-scale of dissipation of turbulent kinetic energy is set by the shear and $\Ast{\epsilon}$ scales inertially. }

\cpc{Therefore, we consider the following scaling (the star is dropped for dimensionless quantities):} 
\begin{equation}
    \Ast{u} = \Ast{U} u, \; \Ast{z} = \Ast{L} z, \; \Ast{t} = \frac{\Ast{L}}{\Ast{U}}t, \; \Ast{S} = \frac{\Ast{U}}{\Ast{L}}S, \; \Ast{N}^{2} = {N_{c}^{\ast}}^{2}N^{2}, \; \Ast{\epsilon} = \frac{U^{3}}{L}\epsilon,  
\end{equation}
where the relevant time-scale has been set by the shear, $N_{c}^{\ast}$ is a typical value of the buoyancy frequency so that $N^{2} = \mathcal{O}(1)$ and, since we are assuming that $\Ast{\epsilon}$ is large enough to sustain an inertial subrange and that the inertial scaling holds, $\epsilon = \mathcal{O}(1)$ (we will assume in the following that $\Ast{\epsilon}$ is constant and therefore consider $\epsilon = 1$ precisely; \rev{in fact, we will show in section~\ref{section:dependence_on_the_parameters} that the precise value of $\epsilon$ does not affect our results}). In practice, ${N_{c}^{\ast}}^{2}$ and $\Ast{U}/\Ast{L}$ are the background stratification and shear of the disturbed profiles considered in the linear stability analysis (section~\ref{section:linear_stability_analysis}).  
System~\eqref{eq:coupled_system_flux_coefficient} then becomes:
\begin{equation}
    \begin{cases}
        \partial_{t} N^2 = \frac{1}{\text{Pr}\text{Re}}\partial_{zz}N^2 + \frac{1}{\text{Ri}_{b}}\partial_{zz}[\Gamma(\text{Ri}_{b}\text{Ri})\epsilon], \\
        \partial_{t} S = \frac{1}{\text{Re}}\partial_{zz}S + \text{Pr}_{T}\partial_{zz}\left[\frac{\Gamma(\text{Ri}_{b}\text{Ri})\epsilon}{S\text{Ri}_{b}\text{Ri}}\right], \\
        \text{Pr} := \frac{\nu^\ast}{ \kappa^\ast}, \ 
        \text{Re} := \frac{\Ast{U}\Ast{L}}{\nu^\ast}, \ 
        \text{Ri}_{b} := \frac{{N_{c}^{\ast}}^{2} \Ast{L}^2}{\Ast{U}^{2}} ,
    \end{cases}
    \label{eq:system_adim}
\end{equation}
where the dependence on three dimensionless parameters (the molecular Prandtl number $\text{Pr}$, the Reynolds number $\text{Re}$ and the bulk Richardson number $\text{Ri}_b$) is made explicit. These two equations are coupled through \rev{the scaled gradient Richardson number} $\text{Ri} := N^{2} / S^{2}$ (always multiplied by $\text{Ri}_{b}$). We expect \rev{staircase formation} to be favoured at larger $\text{Pr}$~\citep{taylor_zhou_2017}. For $\text{Pr} = \mathcal{O}(1)$ we can expect density \rev{staircases} to be smoothed by diffusion, at least for sufficiently small $\text{Re}$ (i.e. sufficiently small Péclet number $\text{Pe} := \text{Pr}\text{Re}$). \rev{Note that the different dimensionless parameters are considered as free parameters independent of each other and of the dynamical quantities. Indeed, the goal of our study is to explore the full parameter space in order to determine regions that are prone to staircase formation but not to assess whether the entire parameter space is actually physically accessible. Indeed, constraining relationships between the different dimensionless parameters would restrict the range of accessible parameters but would not change the stability results presented here. As mentioned previously, we  also assume $\epsilon$ to be constant. Hence, we are focusing our attention on turbulent patches which are relatively homogeneous (in space) and sustained (in time). In practice, we  consider $\epsilon = 1$ but  show in section~\ref{section:dependence_on_the_parameters} that the precise value of $\epsilon$ does not affect our results. Hence the `strength' of the turbulence does not play a major role in our analysis, as soon as this turbulence follows one of the described scalings. }

\subsubsection{Intermediate scaling for moderately stratified flows} 
\label{section:other_possible_scalings}

Instead of considering the inertial scaling introduced in the previous section, we can alternatively assume that the dissipation rate of turbulent kinetic energy $\Ast{\epsilon}$ scales as ${U^{\ast}}^{2}N_{c}^{\ast}$ (with the notation of section~\ref{section:inertial_scaling}). This scaling is relevant, for instance, to \cpc{moderately or strongly stratified flows in the sense that $\text{Fr}_{T} \lesssim 1$ and therefore  the turbulent kinetic energy \cpc{largely} dissipates within a buoyancy time scale and hence $\Ast{\epsilon} \sim {U^{\ast}}^{2}N_{c}^{\ast}$~\citep{garanaik2019inference}. Considering that the \cpc{flow is moderately stratified in an intermediate flow regime, in the sense that the} \cpc{dominant} time scale is \cpc{still} set by the shear (assuming for instance that we are still in a shear-dominated regime \cpc{and so} the shear time scale $T_{S}^{\ast}$ is sufficiently small  compared to the turbulent time scale $T_{T}^{\ast}$ and the buoyancy time scale $T_{b}^{\ast}$~\citep{mater2014unifying}), the system~\eqref{eq:coupled_system_flux_coefficient} becomes:} 
\begin{equation}
    \begin{cases}
        \partial_{t} N^2 = \frac{1}{\text{Pr}\text{Re}}\partial_{zz}N^2 + \frac{1}{\sqrt{\text{Ri}_{b}}}\partial_{zz}[\Gamma(\text{Ri}_{b}\text{Ri})\epsilon], \\
        \partial_{t} S = \frac{1}{\text{Re}}\partial_{zz}S + \text{Pr}_{T}\sqrt{\text{Ri}_{b}}\partial_{zz}\left[\frac{\Gamma(\text{Ri}_{b}\text{Ri})\epsilon}{S\text{Ri}_{b}\text{Ri}}\right]. 
    \end{cases}
    \label{eq:system_other_adim_hyperdiffusion}
\end{equation}
This system is equivalent to the one derived using the inertial scaling (system~\eqref{eq:system_adim}) with the mapping $\sqrt{\text{Ri}_{b}}\Gamma(\text{Ri}_{b}\text{Ri}) \rightarrow \Gamma(\text{Ri}_{b}\text{Ri})$. We will discuss the implications of this intermediate scaling below. 

\subsubsection{Strongly stratified scaling}
\label{section:other_possible_scalings_strong}

For sufficiently strongly stratified flows, consistently with the strong stratification scaling derived by~\citep{garanaik2019inference} and the buoyancy-dominated regime analysed by~\citep{mater2014unifying} for $\text{Fr}_{T} \lesssim 1$ (leading to $\Ast{\epsilon} \sim \Ast{U}^{2}N_{c}^{\ast}$), we can also assume that the time scale is set by the buoyancy (i.e. $\Ast{t} = \frac{1}{N_{c}^{\ast}}t$, assuming for instance $T_{b}^{\ast} \ll T_{S}^{\ast}$) and obtain: 
\begin{equation}
    \begin{cases}
        \sqrt{\text{Ri}_{b}}\partial_{t} N^2 = \frac{1}{\text{Pr}\text{Re}}\partial_{zz}N^2 + \frac{1}{\sqrt{\text{Ri}_{b}}}\partial_{zz}[\Gamma(\text{Ri}_{b}\text{Ri})\epsilon], \\
        \sqrt{\text{Ri}_{b}}\partial_{t} S = \frac{1}{\text{Re}}\partial_{zz}S + \text{Pr}_{T}\sqrt{\text{Ri}_{b}}\partial_{zz}\left[\frac{\Gamma(\text{Ri}_{b}\text{Ri})\epsilon}{S\text{Ri}_{b}\text{Ri}}\right]. 
    \end{cases}
    \label{eq:system_other_adim_hyperdiffusion_buoyancy_time_scale}
\end{equation}
Once again this system is equivalent to system~\eqref{eq:system_adim} with the mappings $\sqrt{\text{Ri}_{b}}\Gamma(\text{Ri}_{b}\text{Ri}) \rightarrow \Gamma(\text{Ri}_{b}\text{Ri})$ and $\sqrt{\text{Ri}_{b}}t \rightarrow t$, and we will also discuss the implications of this strongly stratified scaling below. 

\begin{figure}
    \centerline{
    \includegraphics[width=0.8\linewidth, trim={0 0 0 0}]{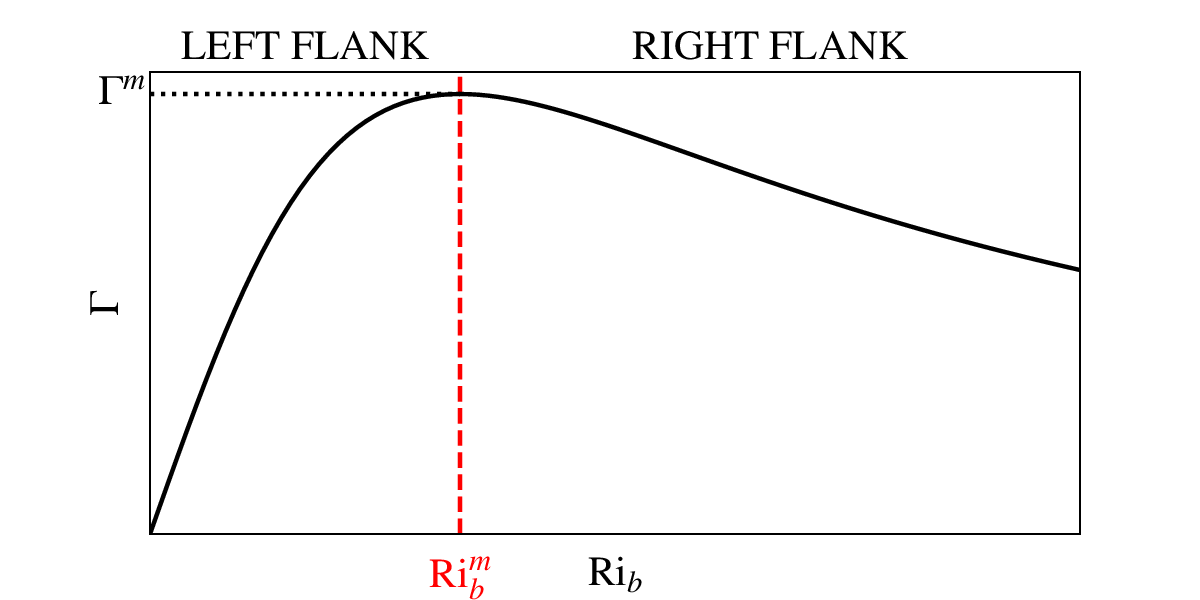}}
    \caption{Schematic representation of the parameterisation of the turbulent flux coefficient $\Gamma$ used throughout this paper. The vertical red dashed line corresponds to the bulk Richardson number that maximizes $\Gamma$. It separates the increasing `left flank' of the $\Gamma$-curve, where $\Gamma$ is an increasing function of $\text{Ri}_{b}$ and the decreasing `right flank' where $\Gamma$ is a decreasing function of $\text{Ri}_{b}$. The horizontal dotted line corresponds to the maximum value of $\Gamma$ (denoted $\Gamma^{m}$). }
    \label{fig:parameterisation_Gamma}
\end{figure}

\subsection{Choice of parameterisation for the flux coefficient}

We must now choose a specific functional form for the parameterisation of the flux coefficient $\Gamma$ \cpc{in terms of} the bulk Richardson number $\text{Ri}_{b}$. Experimental and numerical data \citep{Linden_1979, turner_1968, TheRelationshipbetweenFluxCoefficientandEntrainmentRatioinDensityCurrents} suggest that $\Gamma$ is a non-monotonic function of $\text{Ri}_{b}$ with $\Gamma(\text{Ri}_{b}) \propto \text{Ri}_{b}$ on the increasing flank of $\Gamma$ and $\Gamma(\text{Ri}_{b}) \propto 1/\text{Ri}_{b}^{p}$ (with $p \geq 0$) on the decreasing flank. These scaling regimes \cpc{may be respected with the functional form: } 
\begin{equation}
    \Gamma(\text{Ri}_{b}) = A\frac{\text{Ri}_{b}}{1 + B\text{Ri}_{b}^{p+1}}, 
    \label{eq:parameterisation_gamma}
\end{equation}
where $A$ and $B$ are chosen so that the maximum value of $\Gamma$, attained when $\text{Ri}_{b} = \text{Ri}_{b}^{m} \simeq 1$, is $\Gamma^{m} \simeq 0.2-0.3$ \citep{OntheNatureofTurbulenceinaStratifiedFluidPartITheEnergeticsofMixing}. (As we discuss further below, the specific chosen values of $\Gamma^{m}$ and $\text{Ri}_{b}^{m}$ do not affect the qualitative results presented in this work.) Common values for $p$ are $p=1/2$ and $p=1$ \citep{turner_1968, Linden_1980}. A schematic representation of the parameterisation used is giving in figure~\ref{fig:parameterisation_Gamma}. It should be noted that in what follows we will try to present results that are as general as possible and do not depend strongly on the precise formulation~\eqref{eq:parameterisation_gamma} of $\Gamma$ but only on \rev{the sign of its derivative} and \cpc{asymptotic rate of decrease as $\text{Ri}_b \rightarrow \infty$}. 

\section{Marginal linear stability}
\label{section:linear_stability_analysis}

\subsection{Formulation}
\label{section:linear_stability_analysis_set_up}

\begin{figure}
    \centerline{
    \includegraphics[width=1.1\linewidth, trim={0 2.5cm 0 2cm}]{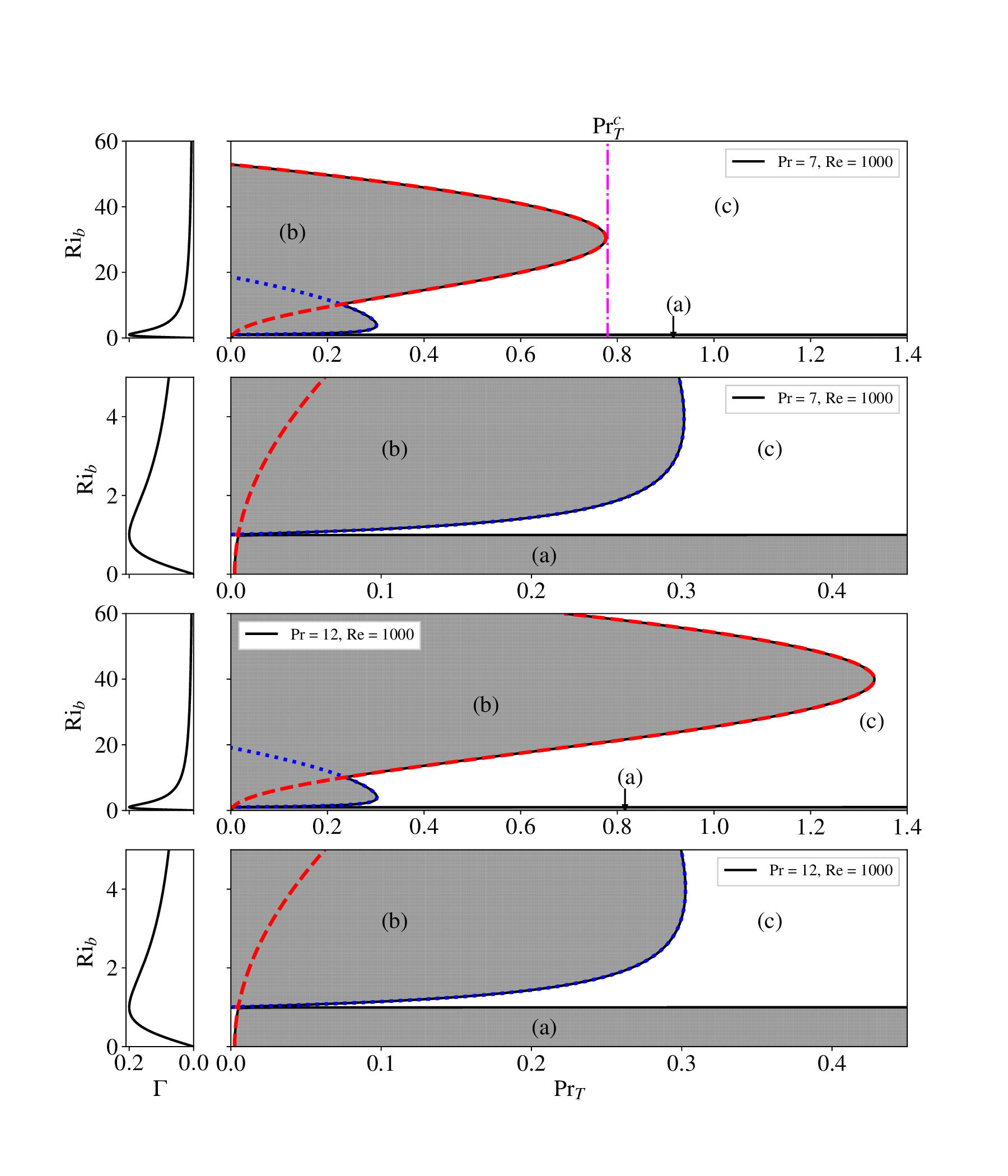}}
    \caption{Range of bulk Richardson numbers $\text{Ri}_{b}$ and turbulent Prandtl numbers $\text{Pr}_{T}$ prone to \rev{staircase} formation for flows with various molecular Prandtl numbers $\text{Pr}$ and $\text{Re} = 1000$, using parameterisation~\eqref{eq:parameterisation_gamma} of $\Gamma$ with $p=1$ (depicted in the left panels). Parameters in region (a) and (b) exhibit staircase formation dynamics (grey shading) whereas parameters in region (c) do not. The horizontal boundary between (a) and (c) corresponds to $\text{Ri}_{b} = \text{Ri}_{b}^{m}$ (i.e. the bulk Richardson number at which $\Gamma$ is maximized). \rev{A zoom of region (a) for the considered values of $\text{Pr}$ is shown the second and fourth rows.} The dotted blue curves correspond to the marginal condition $f=0$ whereas the dashed red curves correspond to the condition $C=0$ \cpc{as defined in \eqref{eq:fcdef}}. The magenta dash-dotted vertical line corresponds to the critical value of the turbulent Prandtl number $\text{Pr}_{T}^{c}$ above which no instability is possible on the decreasing right flank of the $\Gamma$-curve. } 
    \label{fig:stability_plot_fixed_Re_various_Pr_m}
\end{figure}

To investigate the conditions that can support the formation of staircases \rev{starting from linear velocity and buoyancy profiles}, we linearise the system~\eqref{eq:system_adim} around linear profiles of buoyancy and velocity with constant shear $\Ast{U}/\Ast{L}$ and buoyancy frequency $N_{c}^{\ast}$. We therefore assume that the (dimensionless) shear and stratification fields can be decomposed as follows: 
\begin{equation}
    S = 1 + \Tilde{S}, \; N^{2} = 1 + \Tilde{N}^{2}, 
\end{equation}
where the perturbations $\Tilde{S} \ll 1$ and $\Tilde{N}^{2} \ll 1$. Then, at first order in $\Tilde{N}^{2}$ and $\Tilde{S}$:  
\begin{equation}
    \text{Ri} = \frac{1 + \Tilde{N}^{2}}{(1 + \Tilde{S})^{2}} = \left[1 + \Tilde{N}^{2}\right]\left[1 - 2\Tilde{S}\right] = 1 + \Tilde{\text{Ri}}, 
\end{equation}
where $\Tilde{\text{Ri}} = -2\Tilde{S} + \Tilde{N}^{2}$. Considering the dimensionless system~\eqref{eq:system_adim} and considering a constant dissipation rate of turbulent kinetic energy (set to 1, consistently with $\epsilon = \mathcal{O}(1)$ as mentioned above), we obtain, at first order: 
\begin{equation}
    \begin{cases}
    \partial_{t}\Tilde{N}^{2} = \frac{1}{\text{Pr}\text{Re}}\partial_{zz}\Tilde{N}^{2} + \frac{1}{\text{Ri}_{b}}\partial_{zz}[\Tilde{\text{Ri}}\text{Ri}_{b}\Gamma'(\text{Ri}_{b})], \\
    \partial_{t}\Tilde{S} = \frac{1}{\text{Re}}\partial_{zz}\Tilde{S} + \frac{\text{Pr}_{T}}{\text{Ri}_{b}}\partial_{zz}[\Tilde{\text{Ri}}\text{Ri}_{b}\Gamma'(\text{Ri}_{b}) - \Tilde{S}\Gamma(\text{Ri}_{b}) - \Tilde{\text{Ri}}\Gamma(\text{Ri}_{b})], 
    \end{cases}
    \label{eq:linearised_system}
\end{equation}
where we used the first order expansion $\Gamma(\text{Ri}_{b}\text{Ri}) = \Gamma(\text{Ri}_{b}) + \Tilde{\text{Ri}}\text{Ri}_{b}\Gamma'(\text{Ri}_{b})$ with $\Gamma' := \mathrm{d}\Gamma/\mathrm{d}\text{Ri}_{b}$. We now seek normal mode solutions of the form $[\Tilde{S}, \Tilde{N}^{2}] = [\mathcal{A}_{S}, \mathcal{A}_{N}]\text{e}^{\text{i}kz - \text{i}\omega t}$ and obtain a system of linear equations for the eigenvector $[\mathcal{A}_{S}, \mathcal{A}_{N}]$. Since we are interested in non-trivial solutions, we  require the determinant of this system to be zero. This condition is equivalent to the dispersion relation: 
\begin{equation}
    \alpha(k) \omega^{2} - \text{i} \beta(k) \omega + \gamma(k) = 0, 
    \label{eq:dispersion_relation}
\end{equation}
where: 
\begin{equation}
    \begin{cases}
    \alpha(k) = 1, \\ 
    \beta(k) = f(\text{Ri}_{b}, \text{Pr}_{T}, \text{Pr}, \text{Re})k^{2}, \\ 
    \gamma(k) = C(\text{Ri}_{b}, \text{Pr}_{T}, \text{Pr}, \text{Re})k^{4}, 
    \end{cases}
\end{equation}
with: 
\begin{equation}
    \begin{cases}
    f(\text{Ri}_{b}, \text{Pr}_{T}, \text{Pr}, \text{Re}) = (2\text{Pr}_{T} - 1)\Gamma'(\text{Ri}_{b}) - \text{Pr}_{T}\frac{\Gamma(\text{Ri}_{b})}{\text{Ri}_{b}} - \frac{1}{\text{Re}}(1+\frac{1}{\text{Pr}}), \\
    \begin{aligned}
    C(\text{Ri}_{b}, \text{Pr}_{T}, \text{Pr}, \text{Re}) = \text{Pr}_{T}\frac{\Gamma(\text{Ri}_{b})\Gamma'(\text{Ri}_{b})}{\text{Ri}_{b}} + &\frac{\text{Pr}_{T}}{\text{Pr}\text{Re}}\left[-\frac{\Gamma(\text{Ri}_{b})}{\text{Ri}_{b}} + 2\Gamma'(\text{Ri}_{b})\right] \\ - & \frac{\Gamma'(\text{Ri}_{b})}{\text{Re}} - \frac{1}{\text{Pr}\text{Re}^{2}}.
    \end{aligned}
    \end{cases}
    \label{eq:fcdef}
\end{equation}
A wavenumber $k$ is \cpc{thus} unstable if the dispersion relation~\eqref{eq:dispersion_relation} admits a solution \cpc{for frequency $\omega$} with \np{strictly} positive imaginary part. This is equivalent to $\gamma(k) > 0$ or $\gamma(k) \leq 0$ and $\beta(k) > 0$. These conditions are equivalent to $f > 0$ or $C > 0$ and the set of parameters prone to linear instability is therefore: 
\begin{equation}
    \{(\text{Ri}_{b}, \text{Pr}_{T}, \text{Pr}, \text{Re}), f > 0\} \cup \{(\text{Ri}_{b}, \text{Pr}_{T}, \text{Pr}, \text{Re}), C > 0\}. 
\end{equation}
The boundary of this set  separates linearly unstable and stable parameter regions and are plotted in figure~\ref{fig:stability_plot_fixed_Re_various_Pr_m}. 

\subsection{Link with diffusion}
\label{section:link_with_diffusion}

The linearised system~\eqref{eq:linearised_system} can be written in the matrix form: 
\begin{equation}
\renewcommand{\arraystretch}{1.3}
    \left[\begin{array}{c}
         \partial_{t}\Tilde{N}^{2}  \\
         \partial_{t}\Tilde{S} 
    \end{array}\right] 
    = \mathsfbi{D}
    \left[\begin{array}{c}
         \partial_{zz}\Tilde{N}^{2}  \\
         \partial_{zz}\Tilde{S} 
    \end{array}\right],
\end{equation}
where: 
\begin{equation}
    \mathsfbi{D} = \left[\begin{array}{cc}
         \Gamma'(\text{Ri}_{b}) + \frac{1}{\text{Pr}\text{Re}} & -2\Gamma'(\text{Ri}_{b}) \\
         \frac{\text{Pr}_{T}}{\text{Ri}_{b}}[\text{Ri}_{b}\Gamma'(\text{Ri}_{b}) - \Gamma(\text{Ri}_{b})] & \frac{\text{Pr}_{T}}{\text{Ri}_{b}}[-2\text{Ri}_{b}\Gamma'(\text{Ri}_{b}) + \Gamma(\text{Ri}_{b})] + \frac{1}{\text{Re}}
    \end{array}\right]. 
\end{equation}
The matrix $\mathsfbi{D}$ \cpc{may thus be thought of as} a diffusion matrix and the real part of its eigenvalues can be interpreted as effective eddy diffusivities of our problem \rev{(a discussion on the imaginary parts of these eigenvalues is provided in section~\ref{section:regularisation_of_model})}. The trace or this matrix is $-f$ and its determinant is $-C$. Therefore, the instability conditions derived in the previous section are equivalent to the existence of an eigenvalue \cpc{of this matrix} with negative real part and hence an antidiffusive \cpc{dynamical behaviour} that sharpens density gradients. This result can be generalised to the full (nonlinear) system~\eqref{eq:system_adim} (\cpc{as discussed in more detail in} appendix~\ref{appendix:non_linear_diffusion}) but for the purpose of the stability analysis the above (zero-th order) eddy diffusivities suffice to understand the mechanism at hand. 

\subsection{Dependence on the parameters}
\label{section:dependence_on_the_parameters}

\subsubsection{On the increasing left flank of the $\Gamma$-curve}

\begin{figure}
    \centerline{
    \includegraphics[width=\linewidth, trim={0 0 0 0}]{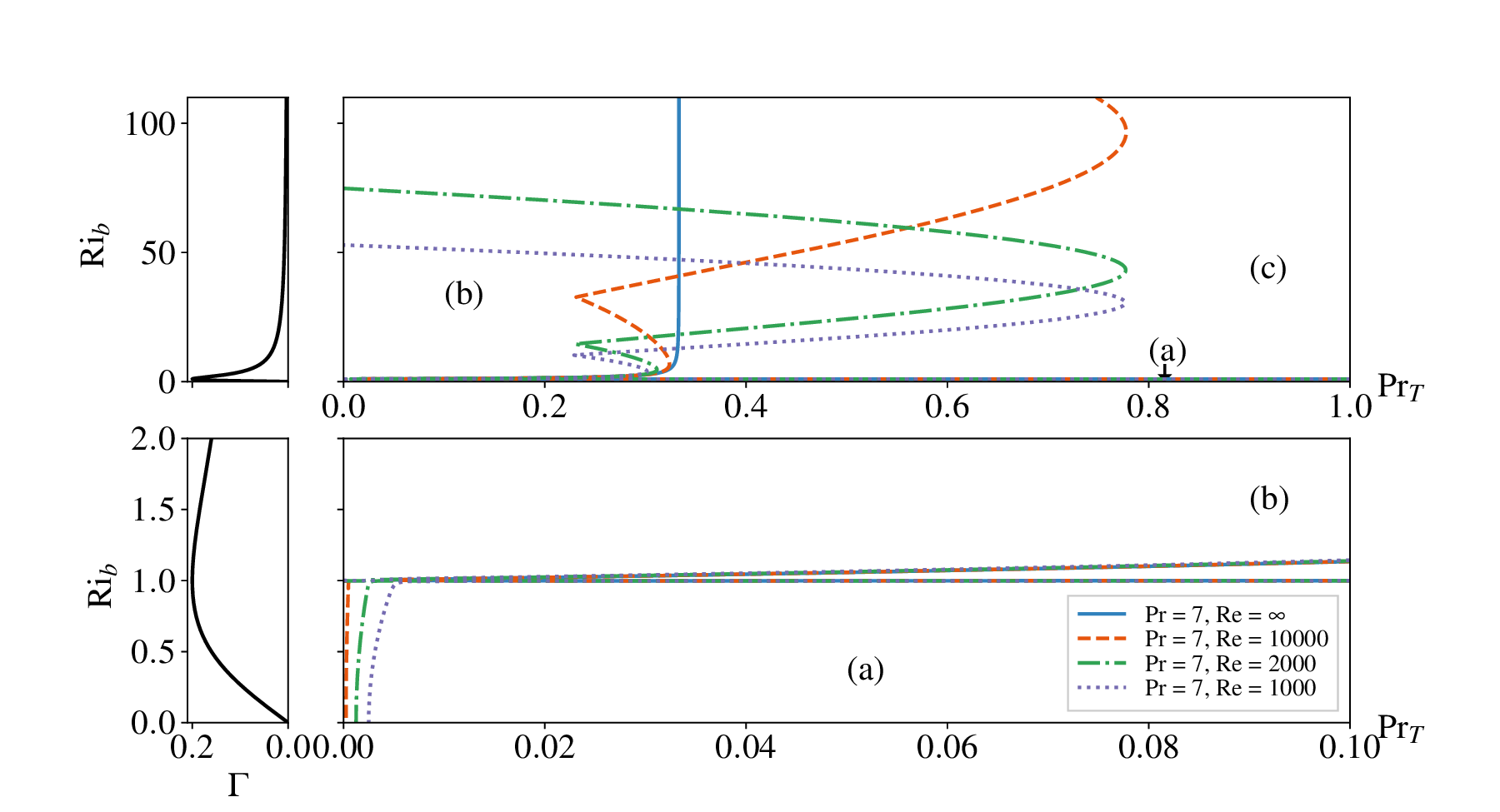}}
    \caption{Range of bulk Richardson numbers $\text{Ri}_{b}$ and turbulent Prandtl numbers $\text{Pr}_{T}$ prone to \rev{staircase} formation for various Reynolds numbers $\text{Re}$ and $\text{Pr} = 7$, using parameterisation~\eqref{eq:parameterisation_gamma} of $\Gamma$ with $p=1$ (depicted in the left panel). \rev{A zoom of region (a) is shown the bottom panel. } For $\text{Re} = \infty$ (i.e. $\Ast{\nu} = 0$) the boundary between (b) and (c) tends towards the vertical line $\text{Pr}_{T} = 1/3$ (see solid blue line). As suggested by the scaling~\eqref{eq:scaling_Reynolds_number}, the critical turbulent Prandtl number $\text{Pr}^c_T$ above which no instability is possible on the decreasing right flank of the $\Gamma$-curve appears to be independent of $\text{Re}$ for $\Ast{\nu} \neq 0$. }
    \label{fig:stability_plot_fixed_Pr_m_various_Re}

    \centerline{
    \includegraphics[width=\linewidth, trim={0 0 0 0}]{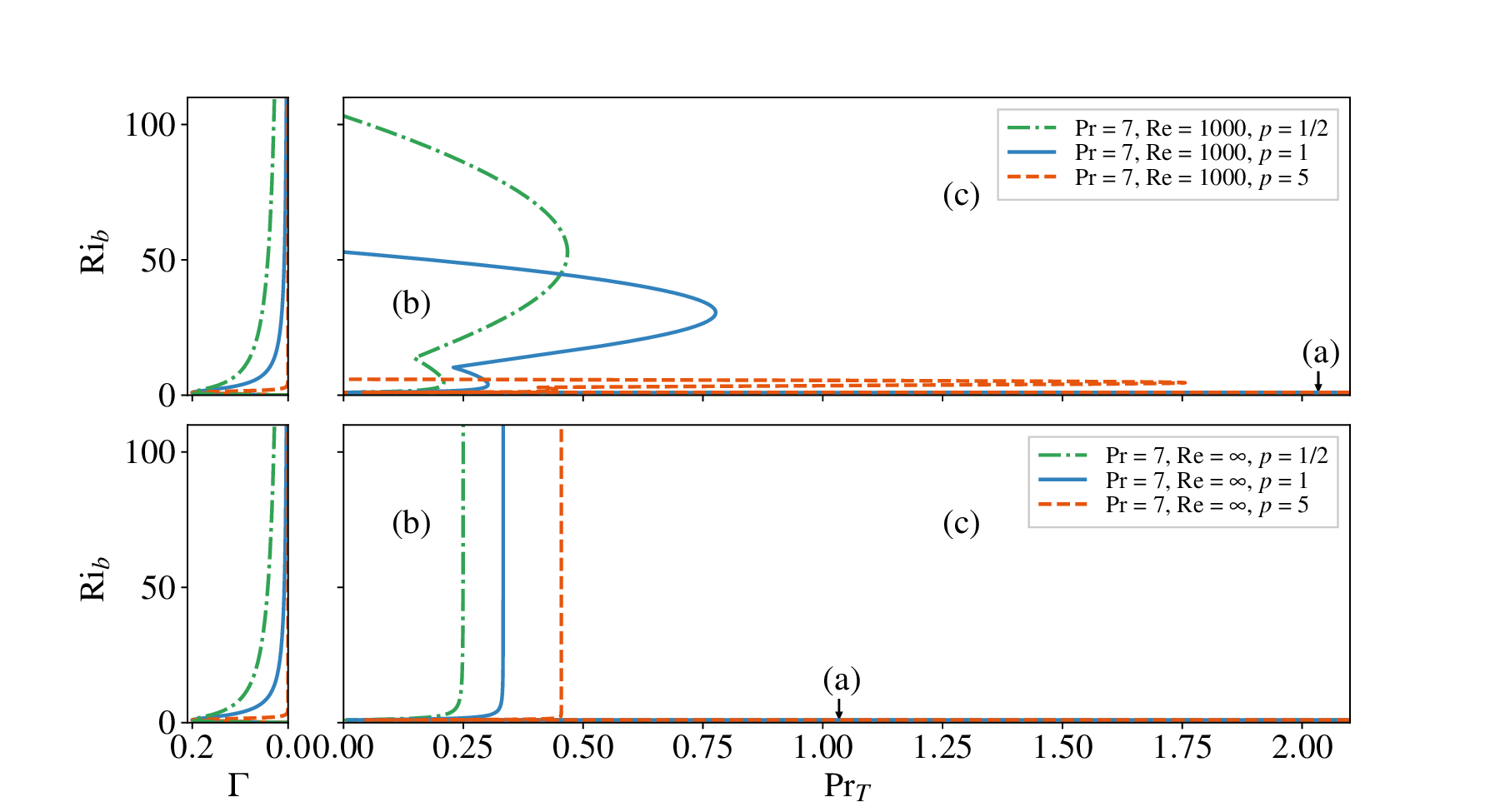}}
    \caption{(Top) Range of bulk Richardson numbers $\text{Ri}_{b}$ and turbulent Prandtl numbers $\text{Pr}_{T}$ prone to \rev{staircase} formation for $\text{Pr} = 7$ and $\text{Re} = 1000$, using parameterisation~\eqref{eq:parameterisation_gamma} of $\Gamma$ with various power laws $p$ (depicted in the left panel). (Bottom) Same with $\text{Re} = \infty$. Note that the behaviour of $\Gamma$ at small $\text{Ri}_{b}$ \cpc{is independent of} $p$. Hence, region (a) is similar to the one depicted in figure~\ref{fig:stability_plot_fixed_Re_various_Pr_m}. } 
    \label{fig:stability_plot_various_m}
\end{figure}

In general, the qualitative stability properties do not depend on the particular functional form of the parameterisation $\Gamma(\text{Ri}_{b})$ but rather on the sign of its derivative $\Gamma'(\text{Ri}_b)$. For sufficiently small $\text{Ri}_b$ such that $\Gamma'(\text{Ri}_b) > 0$, (i.e. on the increasing `left flank' of the flux coefficient curve) the system is linearly unstable for sufficiently large values of $\text{Pr}_T$ (figure~\ref{fig:stability_plot_fixed_Re_various_Pr_m}). As shown in the bottom panel of figure~\ref{fig:stability_plot_fixed_Pr_m_various_Re}, the critical value, denoted $\text{Pr}_{T}^{l}$, can be very small. More precisely, the instability occurs for $\text{Pr}_T > \text{Pr}_{T}^{l} \simeq 0.001$ for flows where $\text{Pr}=7$, $\text{Re}=1000$ and $\Gamma$ increases as $\Gamma(\text{Ri}_{b}) \propto \text{Ri}_{b}$. Moreover, $\text{Pr}_{T}^{l}$ appears to be largely  insensitive to changes in $\text{Pr}$ and tends towards zero as $\text{Re} \rightarrow \infty$ (see bottom panel in figure~\ref{fig:stability_plot_fixed_Pr_m_various_Re}), although it is important to appreciate that the specific 
case $\text{Pr}_{T} = 0$ \np{(that yields $f\leq0$ and $C \leq 0$)} is always linearly stable for flows  on the increasing left flank of the flux coefficient curve.

\begin{figure} 
    \centerline{
    \includegraphics[width=\linewidth, trim={0 0 0 0}]{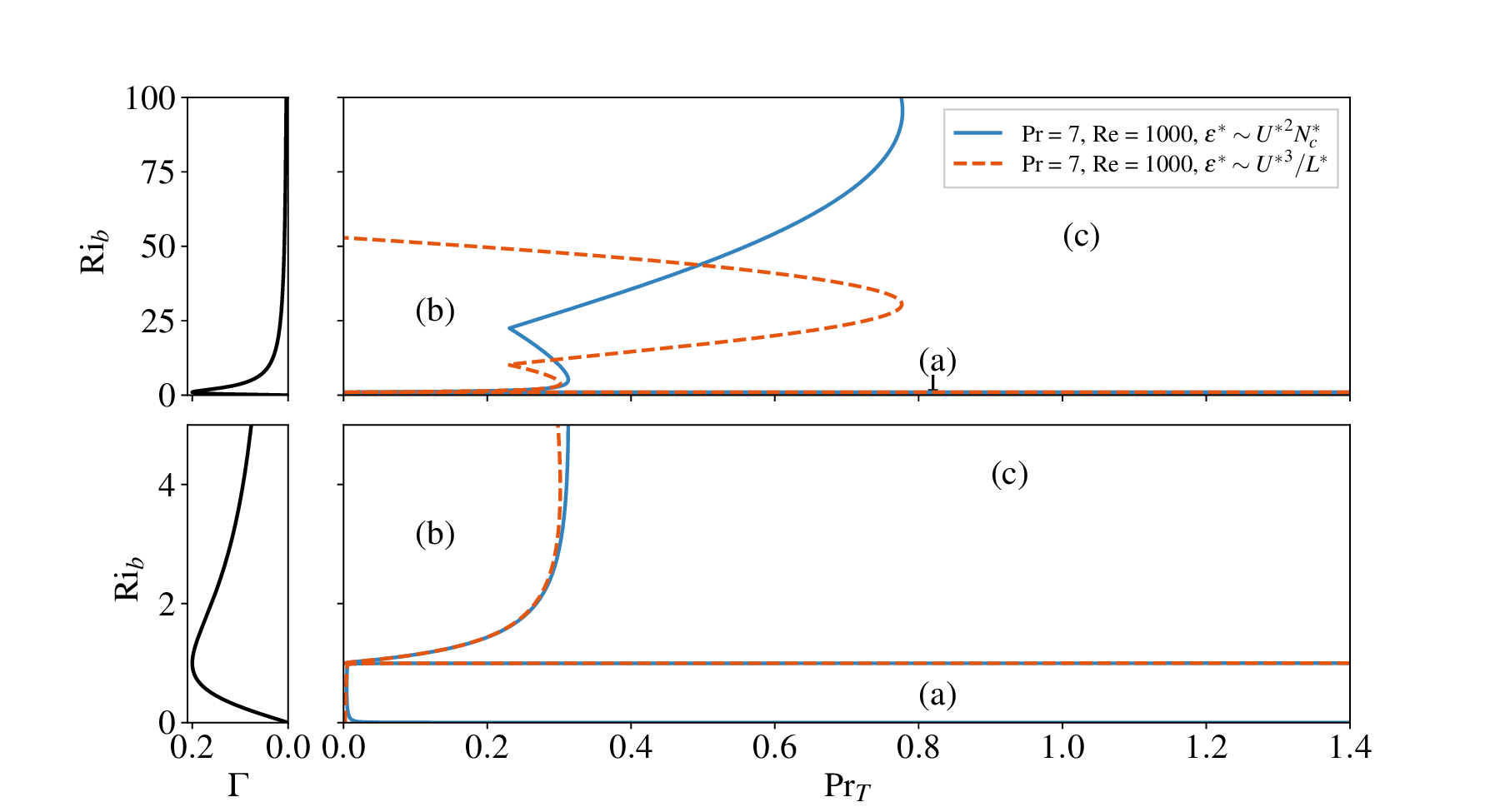}}
    \caption{Range of bulk Richardson numbers $\text{Ri}_{b}$ and turbulent Prandtl numbers $\text{Pr}_{T}$ prone to \rev{staircase} formation for $\text{Pr}=7$ and $\text{Re} = 1000$, using parameterisation~\eqref{eq:parameterisation_gamma} of $\Gamma$ with $p=1$ (depicted in the left panel) for the different scalings for the dissipation rate of turbulent kinetic energy $\Ast{\epsilon}$ discussed in section~\ref{section:non_dimensional_system}. \rev{A zoom of section (a) is shown in the bottom panel.} } 
    \label{fig:stability_plot_various_scalings}

    \centering
    \includegraphics[width=0.9\linewidth, trim={0 0 0 -2cm}]{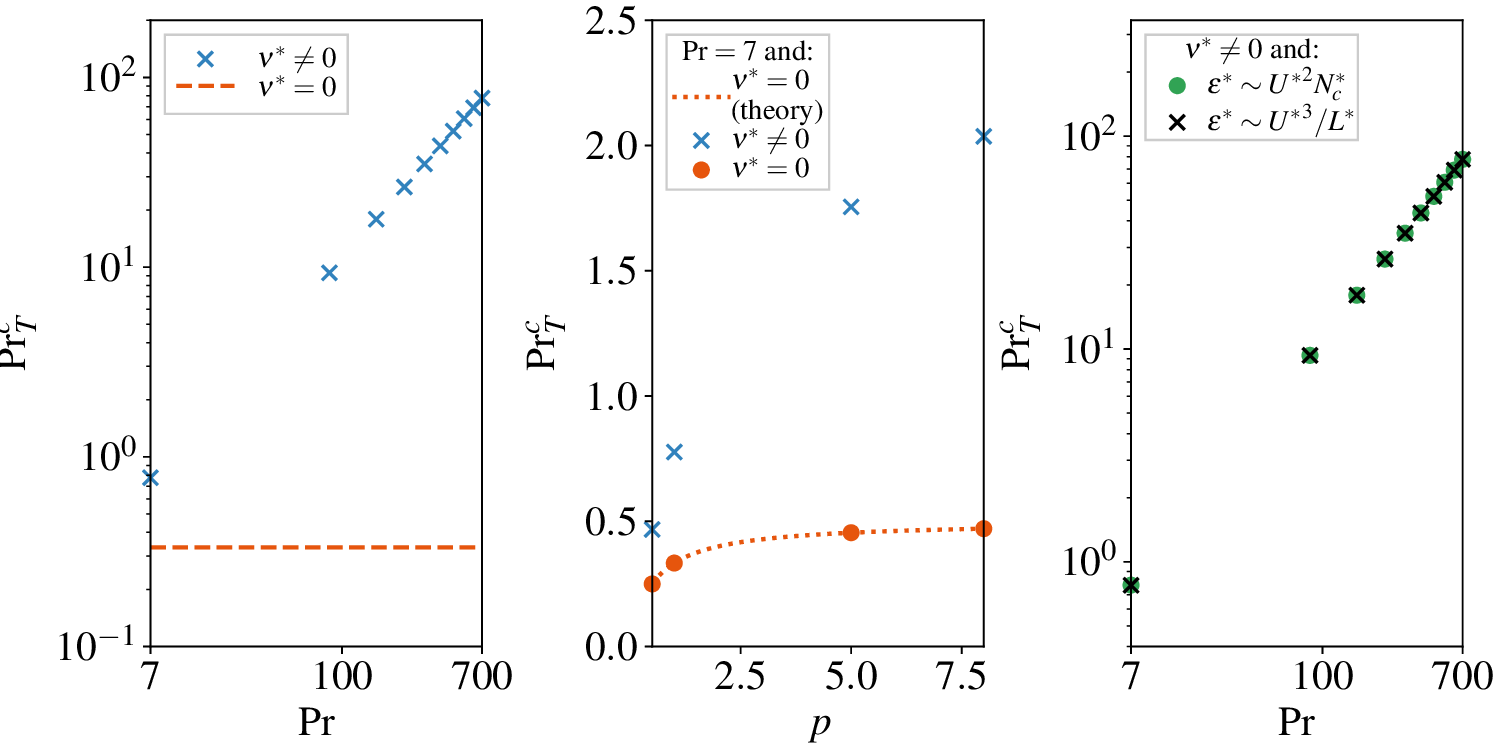}
    \caption{(Left) Critical turbulent Prandtl number $\text{Pr}_T^c$ above which no instability is possible on the decreasing right flank of the $\Gamma$-curve as a function of $\text{Pr}$ for $\nu^\ast \neq 0$ (i.e. finite values of $\text{Re}$) and $\nu^\ast = 0$ (i.e. $\text{Re} \rightarrow \infty$). From the  scaling~\eqref{eq:scaling_Reynolds_number}, for $\nu^\ast \neq 0$, $\text{Pr}^c_T$ does not depend on $\text{Re}$, while for $\nu^\ast = 0$, $\text{Pr}^c_T$ does not depend on $\text{Pr}$ (as shown by the orange dashed line). 
    (Middle) The critical value $\text{Pr}_T^c$ as a function of $p$ for $\text{Pr} = 7$ and $\nu^\ast \neq 0$ (in this case, $\text{Pr}^c_{T}$ depends on $\text{Pr}$ but not on $\text{Re}$) and $\nu^\ast = 0$. When $\nu^\ast=0$,  $\text{Pr}^c_T=p/(2p+1)$ (orange dotted line).
    (Right) The critical value $\text{Pr}^c_T$ as a function of $\text{Pr}$ for the various scalings discussed in section~\ref{section:non_dimensional_system}. }
    \label{fig:variation_Pr_T_c}
\end{figure}

\subsubsection{On the decreasing right flank of the $\Gamma$-curve}

Conversely, on the decreasing `right' flank where $\Gamma'(\text{Ri}_b) < 0$, the flow is linearly stable for sufficiently large $\text{Pr}_T$ and therefore there exists a critical value of the turbulent Prandtl number $\text{Pr}_{T}$, denoted $\text{Pr}_{T}^{c}$ in the subsequent, above which no instability is possible on the decreasing left flank of the $\Gamma$-curve (see top panel in figure~\ref{fig:stability_plot_fixed_Re_various_Pr_m}). 

For finite values of $\text{Re}$ (i.e. $\Ast{\nu} \neq 0$; we discuss the stricly inviscid limit $\Ast{\nu} = 0$ in section~\ref{section:limit_cases}) and parameterisations of the form \eqref{eq:parameterisation_gamma}, the critical value $\text{Pr}_{T}^{c}$ depends only on the molecular Prandtl number $\text{Pr}$ as well as on the decreasing power law $p$ of $\Gamma$, but not on $\text{Re}$. Indeed, if $\Gamma \propto 1/\text{Ri}_{b}^{p}$ the mapping: 
\begin{equation}
    \text{Ri}_{b} \rightarrow a\text{Ri}_{b}, \; \text{Re} \rightarrow a^{p+1}\text{Re}, \; \text{Pr}_{T} \rightarrow \text{Pr}_{T}, \; \text{Pr} \rightarrow \text{Pr},   
    \label{eq:scaling_Reynolds_number}
\end{equation}
maps $f \rightarrow 1/a^{p+1}f$ and $C \rightarrow 1/a^{2p+2}C$, and so crucially does not affect the sign of these functions (and hence the associated stability properties). Hence, changing $\text{Re}$ only stretches the boundary between linearly unstable and stable regions in the $\text{Ri}_{b}$ direction, as depicted in figure~\ref{fig:stability_plot_fixed_Pr_m_various_Re}, and do not affect $\text{Pr}_{T}^{c}$. Similarly, variations of the parameter $A$ in~\eqref{eq:parameterisation_gamma} does not significantly affect $\text{Pr}_{T}^{c}$. This can be established through consideration of the mapping:
\begin{equation}
    \Gamma \rightarrow a\Gamma, \text{Ri}_{b} \rightarrow \text{Ri}_{b}, \; \text{Re} \rightarrow \frac{1}{a}\text{Re}, \; \text{Pr}_{T} \rightarrow \text{Pr}_{T}, \; \text{Pr} \rightarrow \text{Pr},  
    \label{eq:scaling_for_A}
\end{equation}
which maps $f \rightarrow af$ and $C \rightarrow a^{2}C$ which once again does not affect the sign of $f$ and $C$, key to the stability properties. Moreover, we have seen previously that scaling $\text{Re}$ is equivalent to stretching the marginal stability curves in the $\text{Ri}_{b}$ direction only. Therefore, the critical value $\text{Pr}_{T}^{c}$ is unaffected by changes of $A$. \rev{Note that using a similar mapping, we can show that the choice of $\epsilon$ in~\eqref{eq:system_adim} does not affect $\text{Pr}_{T}^{c}$. Indeed, this constant only comes into play when multiplied by $\Gamma$.} Likewise, the parameter $B$ in~\eqref{eq:parameterisation_gamma} does not affect $\text{Pr}_{T}^{c}$. More precisely, variations in $B$ translate the marginal stability curves in the $\text{Ri}_{b}$ direction (since this parameter only affects the value $\text{Ri}_{b}^{m}$ of the bulk Richardson number that maximizes $\Gamma$). As a result, the critical value $\text{Pr}_{T}^{c}$ depends on the decreasing power law $p$ but not on the particular choices for $A$ and $B$ in~\eqref{eq:parameterisation_gamma}, suggesting some robustness of our results with respect to the parameterisation of the flux coefficient. 

Variations of $\text{Pr}_{T}^{c}$ with $\text{Pr}$ are depicted on the left panel of figure~\ref{fig:variation_Pr_T_c}. The critical value $\text{Pr}^c_T$ increases with $\text{Pr}$, consistently with the fact that \rev{staircase formation} is favoured at large molecular Prandtl number~\citep{taylor_zhou_2017}. More precisely, for $p=1$ and $\text{Pr} = 7$ (the typical value of $\text{Pr}$ for thermally-stratified water),  $\text{Pr}^c_T \simeq 0.8$ whereas for $\text{Pr} = 700$ (i.e. water where density is set by salinity),  $\text{Pr}^c_T \simeq 80$. Variation of the critical value of $\text{Pr}_{T}^{c}$ with $p$ are depicted on the middle panel of figure~\ref{fig:variation_Pr_T_c}. For example, for $\text{Pr} = 7$ (and $\Ast{\nu} \neq 0$), the critical value increases from $\text{Pr}^c_T \simeq 0.5$ when $p=1/2$ to $\text{Pr}^c_T \simeq 2$ for $p=8$. 

All in all, for $\text{Pr} = 7$ and $p$ of order unity, the critical value of the turbulent Prandtl number is found to be around $\text{Pr}^c_T \simeq 0.5 - 0.8$. Importantly, this key result concerning the critical turbulent Prandtl number does not depend on the scalings for the dissipation rate of turbulent kinetic energy \rev{considered in this paper}. Indeed, for the intermediate scaling presented in section \ref{section:other_possible_scalings} leading to system 
\eqref{eq:system_other_adim_hyperdiffusion}, the associated
mapping does not change $\text{Pr}_T^c$, but rather only stretches the marginal statibility curves in the $\text{Ri}_{b}$ direction as shown in figure~\ref{fig:stability_plot_various_scalings} and~\ref{fig:variation_Pr_T_c}. Analogously, for the strongly stratified scaling presented in section \ref{section:other_possible_scalings_strong} leading to system \eqref{eq:system_other_adim_hyperdiffusion_buoyancy_time_scale}, the associated mapping again does not change $\text{Pr}_T^c$, but rather  stretches the marginal stability curves in the $\text{Ri}_b$ direction 
and modifies the magnitude of the (unstable) growth rates. 

Note that if $\Gamma$ saturates at a constant value instead of monotonically decreasing towards zero at large bulk Richardson numbers, then $\Gamma'(\text{Ri}_{b}) = 0$ for $\text{Ri}_{b}$ large enough and both $f$ and $C$ become negative. Then the system is linearly stable for all $\text{Pr}_{T}$ and $\text{Pr}_{T}^{c} = 0$. 

We can also define a \cpc{critical bulk Richardson number $\text{Ri}^c_b$ above which no instability is possible. For parameterisations with $\Gamma \propto 1/\text{Ri}_{b}^{p}$,  $\text{Ri}_b^c = {\cal O} (\text{Pr}\text{Re})^{\frac{1}{p+1}}$ under the assumption that the dissipation rate exhibits inertial scaling.
When the dissipation rate exhibits the intermediate and strongly stratified scalings presented in sections \ref{section:other_possible_scalings}
and \ref{section:other_possible_scalings_strong}, $\text{Ri}_b^c = {\cal O}(\text{Pr}\text{Re})^{\frac{1}{p+1/2}}$. The fact that this limit increases with  $\text{Re}$ seems reasonable (as viscous effects are expected to inhibit perturbation growth), while  the fact that $\text{Ri}_b^c$  increases with $\text{Pr}$ is consistent with previous studies establishing that \rev{staircase formation} seems to be favoured for large molecular Prandtl numbers \citep{taylor_zhou_2017}. }

\subsection{Limit cases}
\label{section:limit_cases}

In this section we analyse four limits of our problem, namely $\nu^\ast = 0$, ${N_{c}^{\ast}}^{2} = 0$, $\text{Pr}_{T} = 0$ and the case $\text{Pr} \ll 1$, \np{$\text{Pr}\text{Re} \ll 1$}. 

\subsubsection{Case $\nu^\ast = 0$}
\np{We first consider the inviscid limit $\nu^\ast = 0$ (i.e. $\text{Re} \rightarrow \infty$). On the increasing left flank of the $\Gamma$-curve, we have $C > 0$ for all $\text{Pr}_{T} > 0$ and therefore the system is unstable in this case (the case $\text{Pr}_{T} = 0$ gives $C=0$ and $f<0$ and is therefore stable). Conversely,} on the decreasing right flank of the $\Gamma$-curve, the condition $C \geq 0$ \cpc{is no longer well-defined, as is apparent from the definition \eqref{eq:fcdef}}. Moreover, the condition $f \geq 0$ cannot be satisfied on the decreasing \cpc{right flank of the $\Gamma$-curve} for $\text{Pr}_{T} \geq 1/2$. Hence, on the decreasing \cpc{right flank of the $\Gamma$-curve} and for $\Ast{\nu} = 0$, if $\text{Pr}_{T} \geq 1/2$ the system is linearly stable and this result is valid for any decreasing $\Gamma$-curve, \cpc{a  result first derived by \citep{kranenburg1980stability}}. For $\Gamma \propto 1/\text{Ri}_{b}^{p}$, this bound can be sharpened to $\text{Pr}_{T} \geq p / (2p + 1)$, as shown in figure~\ref{fig:stability_plot_various_m} and~\ref{fig:variation_Pr_T_c}. (This result is not in contradiction with the \cpc{critical value $\text{Pr}_T^c$} for instability found in section~\ref{section:dependence_on_the_parameters} using the scaling~\eqref{eq:scaling_Reynolds_number} since this scaling is valid for finite values of the Reynolds number $\text{Re}$ only.)

\subsubsection{Case ${N_{c}^{\ast}}^{2} = 0$}
\label{section:N2_0}

In the unstratified limit  ${N_{c}^{\ast}}^{2} = 0$ there is \cpc{(of course)} no buoyancy to mix. The above analysis suggests that the case ${N_{c}^{\ast}}^{2} = 0$ (which is equivalent to $\text{Ri}_{b} = 0$) is linearly unstable (at least for large enough turbulent Prandtl numbers) and we therefore expect instabilities to develop in the velocity field rather than in the buoyancy field.  Since the scalings presented in section~\ref{section:non_dimensional_system} cannot be used when ${N_{c}^{\ast}}^{2} = 0$, we consider here the dimensional system~\eqref{eq:coupled_system_flux_coefficient}. Let us first linearise the system~\eqref{eq:coupled_system_flux_coefficient} around a state of zero stratification, i.e.  we decompose the buoyancy field (in dimensional form) as $\Ast{N}^{2} = 0 + \Tilde{\Ast{N}}^{2}$ (where $\Tilde{\Ast{N}}^{2}$ is a small perturbation), implying the decomposition $\text{Ri}_{g} = 0 + \Tilde{\Ast{N}}^{2} / \Ast{S}^{2}$ for the Richardson number (no assumptions are made about the size of $\Ast{S}$). We then obtain, at first order, \rev{in dimensional form}: 
\begin{equation}
    \begin{cases}
        \partial_{\Ast{t}}\Tilde{\Ast{N}}^{2} = \kappa^\ast\partial_{\Ast{z}}^{2}\Tilde{\Ast{N}}^{2} + \Gamma'(0)\partial_{\Ast{z}}^{2}\left[\Ast{\epsilon}\frac{\Tilde{\Ast{N}}^{2}}{\Ast{S}^{2}}\right], \\
        \partial_{\Ast{t}}\Ast{S} = \nu^\ast\partial_{\Ast{z}}^{2}\Ast{S} +  \text{Pr}_{T}\Gamma'(0)\partial_{\Ast{z}}^{2}[\frac{\Ast{\epsilon}}{\Ast{S}}]. 
    \end{cases}
\end{equation}
The $\Tilde{\Ast{N}}^{2}$-equation is parabolic, and using a maximum principle (assuming, for example, Dirichlet boundary conditions), we can show that the perturbation $\Tilde{\Ast{N}}^{2}$ will remain at all times bounded by the initial perturbation $\max_{\Ast{z}}\lvert \Tilde{\Ast{N}}^{2}(\Ast{t}=0, \Ast{z})\rvert$. Therefore, starting from a perturbation of the buoyancy profile small enough such that the above linearisation stands, this perturbation will remain small and any interesting dynamics will develop in the velocity profile alone. 

\subsubsection{Case $\mathrm{Pr}_{T} = 0$}
\label{section:Pr_T_0}
In the limit of small turbulent Prandtl numbers, any layering dynamics  will occur preferentially in the buoyancy field. More precisely, setting $\text{Pr}_{T} = 0$ (and $\epsilon = 1$ for clarity) in the dimensionless system~\eqref{eq:system_adim} yields:  
\begin{equation}
    \begin{cases}
        \partial_{t} N^2 = \partial_{zz}\left[\frac{1}{\text{Pr}\text{Re}}N^2 + \frac{1}{\text{Ri}_{b}}\Gamma(\text{Ri}_{b}\text{Ri})\right], \\
        \partial_{t} S = \frac{1}{\text{Re}}\partial_{zz}S,  
    \end{cases}
    \label{eq:system_adim_Pr_T_0}
\end{equation}
and the system is now decoupled. The second equation is a purely diffusive equation that will damp any perturbation in the shear profile exponentially fast on molecular time-scales. Hence, the shear $S$ will tend towards the constant profile $S_{0} = 1$, remembering that this system is dimensionless. The $N^2$-equation is prone to the Phillips mechanism, as \rev{staircases} will form in the buoyancy profile when the right-hand side  $F(N^{2}) = \frac{1}{\text{Pr}\text{Re}}N^{2} + \frac{1}{\text{Ri}_{b}}\Gamma\left(\text{Ri}_{b}\frac{N^{2}}{S_{0}}\right)$ is a decreasing function of $N^2$, whereas any perturbation will be damped on the increasing flank of this function. This observation is consistent with linear stability analysis. Indeed, for $\text{Pr}_{T} = 0$ we \cpc{obtain the equivalent condition for instability}: 
\begin{equation}
    f > 0 \text{ or } C > 0 \Leftrightarrow \Gamma'(\text{Ri}_{b}) < -\frac{1}{\text{Pr}\text{Re}}. 
    \label{eq:stability_condition_Pr_T_0}
\end{equation}

Therefore, the case $\text{Pr}_{T} = 0$ is equivalent to the Phillips mechanism as formulated in \citep{phillips1972turbulence} and in this limit the system is linearly stable for $\text{Ri}_{b} \leq \text{Ri}_{b}^{m}$ and \rev{staircase formation} can only happen for sufficiently stratified flows. 
As shown in section~\ref{section:dependence_on_the_parameters}, this result can be extended to $\text{Pr}_{T} \ll 1$. On the contrary, for larger value of $\text{Pr}_{T}$, the instability seems to be favoured for small bulk Richardson numbers i.e. sufficiently weakly stratified flows on the increasing left flank of the $\Gamma$-curve. Hence, once again the central conclusion is that in the presence of shear and buoyancy driven turbulence, the Phillips mechanism for staircase formation in strongly stratified flows seems to survive only in the limit of small turbulent Prandtl numbers. 

\subsubsection{Case $\mathrm{Pr} \ll 1$, $\mathrm{Pr}\mathrm{Re} \ll 1$}

Let us consider the case of small molecular Prandtl and Péclet numbers, where the Péclet number is defined as $\text{Pe} := \text{Pr}\text{Re}$ and can be understood as the ratio of the advective and diffusive time-scales. This case is relevant to astrophysical stratified turbulent flows where $\text{Pr}$ usually ranges between $10^{-9}$ and $10^{-5}$ and can therefore sustain small $\text{Pe}$, high $\text{Re}$ regimes \citep{garaud2015excitation}. In the limit $\text{Pr} \ll 1$ and $\text{Pe} \ll 1$ (and considering finite Reynolds, bulk Richardson and turbulent Prandtl numbers), \cpc{consideration
of \eqref{eq:fcdef} shows that $f \rightarrow -\infty$, while $C \rightarrow \infty$ for $\text{Pr}_{T}\left[-\frac{\Gamma(\text{Ri}_{b})}{\text{Ri}_{b}} + 2\Gamma'(\text{Ri}_{b})\right] - \frac{1}{\text{Re}} > 0$ and $ C\rightarrow -\infty$ otherwise. Therefore on the decreasing right flank of the $\Gamma$-curve (i.e. where $\Gamma'(\text{Ri}_{b}) \leq 0$), both $f$ and $C$ are negative and the system is linearly stable. Hence, for sufficiently large $\text{Ri}_{b}$, \rev{staircase formation} is prohibited, consistently with the fact that buoyancy anomalies are rapidly diffused for $\text{Pe} \ll 1$ \citep{cope_garaud_caulfield_2020}. }

\section{Instability properties}
\label{section:instability_properties}

\subsection{Wavenumber dependence}

\begin{figure}
    \centerline{
    \includegraphics[width=0.7\linewidth, trim={0 1cm 0 1cm}]{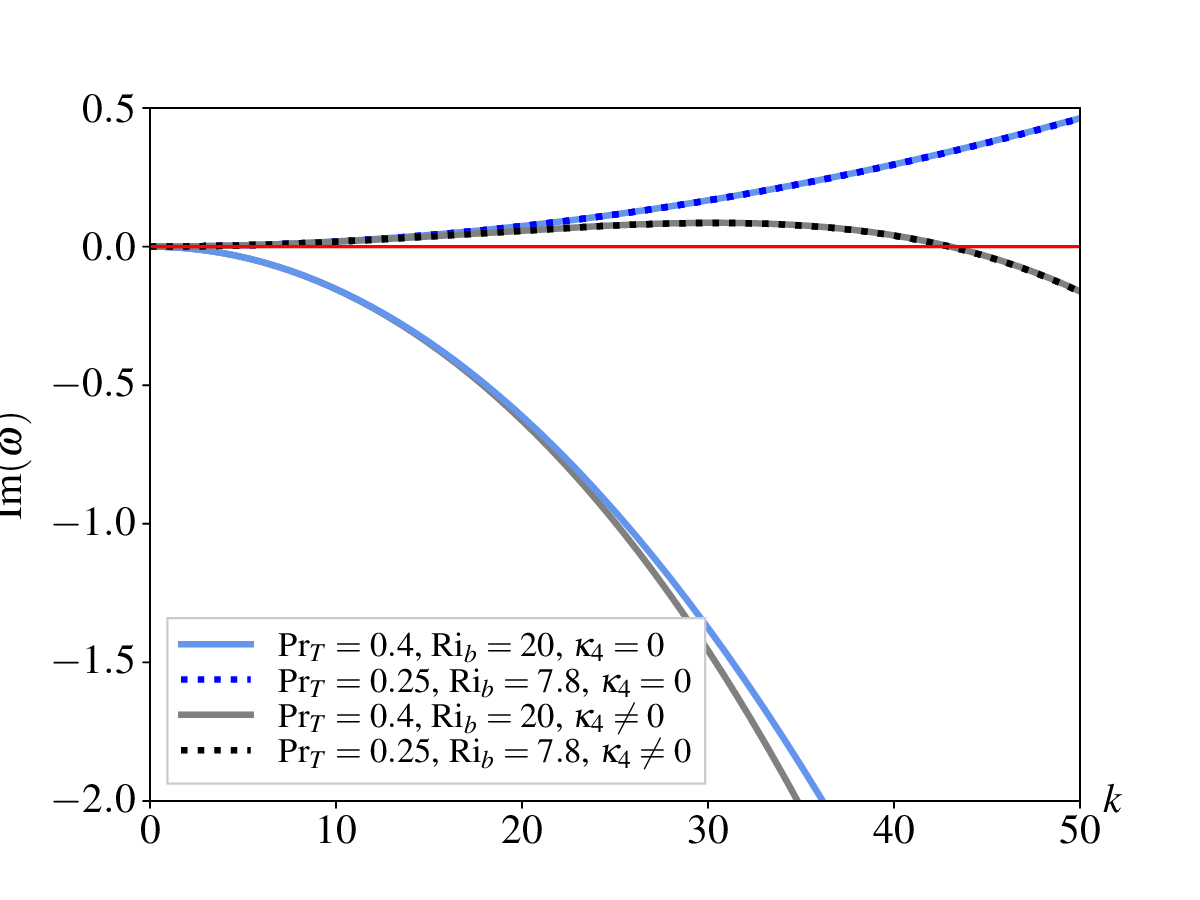}}
    \caption{Growth rate as a function of the wavenumber $k$ for various sets of parameters and with or without hyperdiffusion $\kappa_{4}$. The red horizontal line correspond to $\text{Im}(\omega) = 0$. The blue curves (corresponding to $\kappa_{4} = 0$) unveil an `ultraviolet catastrophe' of unbounded growth at small scales (i.e. large vertical wavenumbers $k$). }
    \label{fig:growth_rate_wave_number_dependence}
\end{figure}

The dispersion relation~\eqref{eq:dispersion_relation} yields: 
\begin{equation}
    \omega(k) = \frac{1}{2}\left[\text{i}k^{2}f \pm \Delta_{0}(k)^{1/2}\right], 
\label{eq:dispersion_relation_no_hyperdiffusion}
\end{equation}
where $\Delta_{0}(k) := (-\text{i}\beta)^{2} - 4\alpha\gamma = (-f^{2} - 4C)k^{4}$. Therefore $\omega \propto k^{2}$ and any perturbation of linearly unstable velocity and buoyancy profiles will grow with growth rates that are proportional to the square of the vertical wavenumber, \cpc{as shown in figure~\ref{fig:growth_rate_wave_number_dependence}}. Hence, the model exhibits an `ultraviolet catastrophe' of unbounded growth at small scales. This unphysical 
behaviour is a consequence of the fact that flux-gradient parameterisations of eddy turbulent fluxes inevitably break down at small scales (namely scales below the representative scale of the turbulent microstructures that shape the flow on larger scales). Similar issues have been encountered in the double-diffusion literature. 
For example,~\citet{radko_2014} shows that fingering flux-gradient models tend to fail `when the size of the phenomenon of interest is comparable to the scale of microstructure which those laws strive to parameterize'.

\cpc{Furthermore,} \citet{Ma_Peltier_2021_Gamma_instability} encounter an ultraviolet catastrophe when studying salt-fingering-engendered thermohaline staircases using a diffusive parameterisation of heat and salinity turbulent fluxes. Again, the problem originates from the assumption that gradient-flux laws are valid at all scales, even those that are smaller than salt-finger widths. To resolve the problem, hyperdiffusive terms were added to the model to correct the model and dampen perturbations at small scales. This can be justified by a multiscale analysis of the problem (as performed by \citet{radko_2019}) that takes into account the interaction between small (microstructure turbulence) and larger scales. 

\subsection{Regularisation of the model at small scales}
\label{section:regularisation_of_model}

Following the ideas of \citet{radko_2019} and \citet{Ma_Peltier_2021_Gamma_instability}, we add hyperdiffusion terms 
to regularise our dimensionless system~\eqref{eq:system_adim} as follows: 
\begin{equation}
    \begin{cases}
        \partial_{t} N^2 = \frac{1}{\text{Pr}\text{Re}}\partial_{zz}N^2 + \frac{1}{\text{Ri}_{b}}\partial_{zz}[\Gamma(\text{Ri}_{b}\text{Ri})\epsilon] - \kappa_{4}\partial_{z}^{4}N^{2}, \\
        \partial_{t} S = \frac{1}{\text{Re}}\partial_{zz}S + \frac{\text{Pr}_{T}}{\text{Ri}_{b}}\partial_{zz}\left[\frac{\Gamma(\text{Ri}_{b}\text{Ri})\epsilon}{S\text{Ri}}\right] - \kappa_{4}\partial_{z}^{4}S, 
    \end{cases}
    \label{eq:system_adim_hyperdiffusion}
\end{equation}
where the scaling factor $\kappa_{4}$ will be chosen later. 

\cpc{Importantly,} the addition of hyperdiffusion does not change \cpc{our}  stability results. \rev{Indeed}, following the same \cpc{approach} as in section~\ref{section:linear_stability_analysis}, it can be shown that the dispersion relation becomes: 
\begin{equation}
    \alpha_{h}(k) \omega^{2} - \text{i} \beta_{h}(k) \omega + \gamma_{h}(k) = 0, 
    \label{eq:dispersion_relation_hyperdiffusion}
\end{equation}
with: 
\begin{equation}
    \begin{cases}
    \alpha_{h}(k) = 1, \\ 
    \beta_{h}(k) = -2\kappa_{4}k^{4} + k^{2}f(\text{Ri}_{b}, \text{Pr}_{T}, \text{Pr}, \text{Re}), \\ 
    \gamma_{h}(k) = k^{4}[-\kappa_{4}^{2}k^{4} + \kappa_{4}f(\text{Ri}_{b}, \text{Pr}_{T}, \text{Pr}, \text{Re})k^{2} + C(\text{Ri}_{b}, \text{Pr}_{T}, \text{Pr}, \text{Re})], 
    \end{cases}
\end{equation}
\cpc{where $f$ and $C$ are identical to the previous 
expressions given in \eqref{eq:fcdef}.} A wavenumber $k$ is unstable if $\gamma_{h}(k) > 0$ or $\gamma_{h}(k) \leq 0$ and $\beta_{h}(k) > 0$. The condition $\beta_{h}(k) > 0$ is equivalent to:  
\begin{equation}
    k^{2} < f / 2\kappa_{4}. 
\end{equation}
Therefore the existence of $k \geq 0$ such that $\beta_{h}(k) > 0$ is equivalent to $f > 0$. Then, if a set of parameters $(\text{Ri}_{b}, \text{Pr}_{T}, \text{Pr}, \text{Re})$ satisfy $f > 0$ we can find small wavenumbers $k < (f/2\kappa_{4})^{1/2}$ that are linearly unstable. \rev{Regarding the condition on $\gamma_{h}$, we can show} using the fact that $\gamma_{h}/k^{4}$ is a polynomial of degree two in $k^{2}$ that the existence of a wavenumber $k \geq 0$ such that $\gamma_{h}(k) > 0$ is equivalent to $C > 0$ or $C \leq 0$ and $f > 0$ and $f^{2} > -4C$. Combining the above conditions, the unstable set of parameters is defined by $\{f>0\} \cup \{C>0\} \cup [\{C \leq 0\} \cap \{f > 0\} \cap \{f^{2} > -4C\}] = \{f>0\} \cup \{C>0\}$, exactly as in section~\ref{section:linear_stability_analysis}. Moreover, using again the polynomial structure of $\gamma_{h} / k^{4}$, we can show that the maximum wavenumber satisfying $\gamma_{h}(k) \geq 0$ is  ${\cal O}(\kappa_{4}\max(f, \sqrt{C}) / \kappa_{4}^{2})^{1/2} = {\cal O}(\max(f, \sqrt{C}) / \kappa_{4})^{1/2}$ when it exists. \cpc{Therefore, since the magnitude of $f$ and $\sqrt{C}$ is set by $1/\text{Re}$ for the range of Reynolds numbers considered here, the largest unstable wavenumber, if it exists, is of order ${\cal O}(1 / \text{Re}\kappa_{4})^{1/2}$. }

\begin{figure}
    \centerline{
    \includegraphics[width=\linewidth, trim={0 0 0 1cm}]{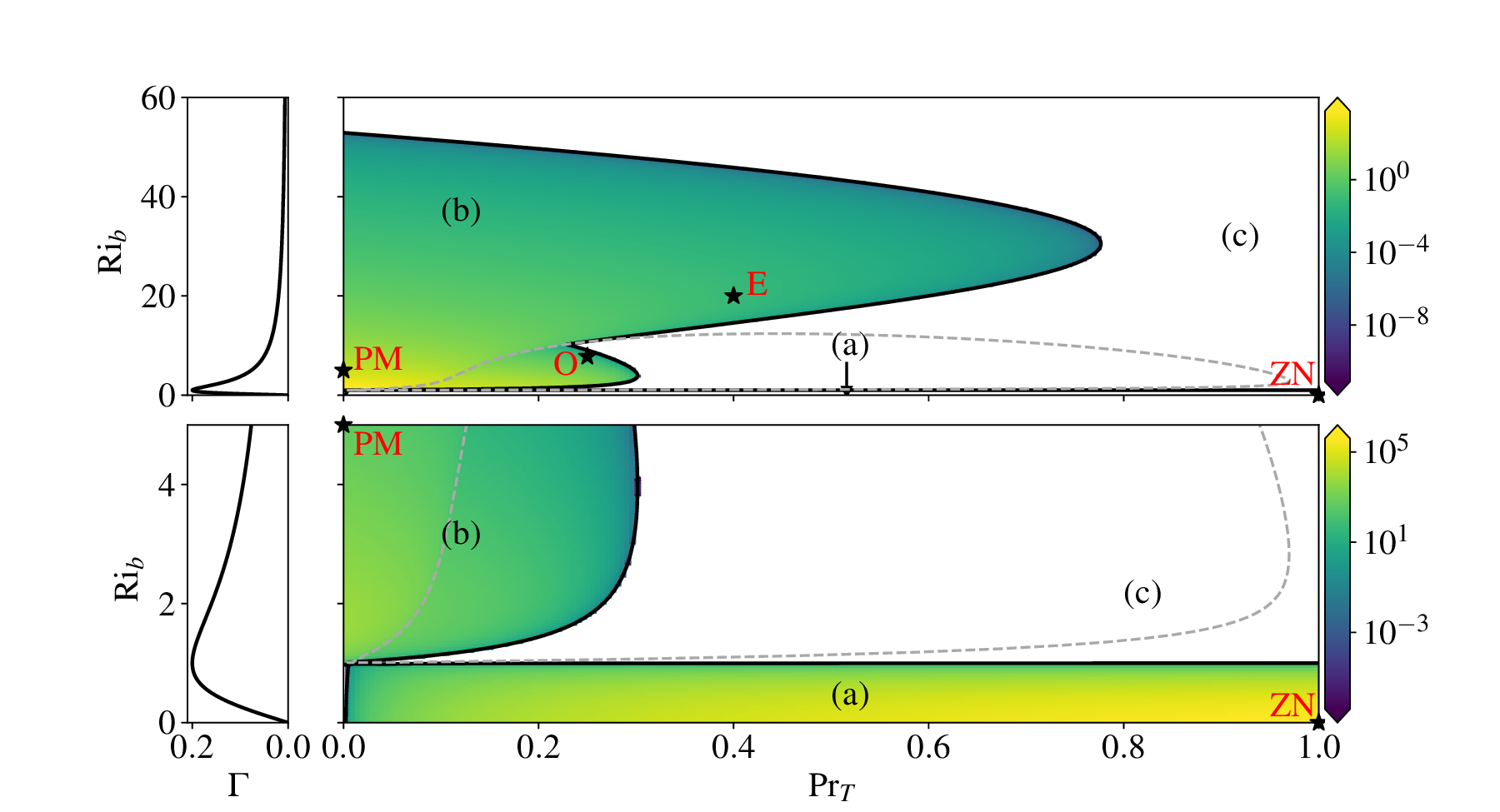}}
    \caption{Variation of maximum growth rate $\sigma_{\text{max}}$ (\cpc{on a} logarithmic scale) with bulk Richardson number $\text{Ri}_{b}$ and  turbulent Prandtl number $\text{Pr}_{T}$ for $\text{Pr} = 7$, $\text{Re} = 1000$, $\kappa_{4} = 10^{-7}$ and using parameterisation~\eqref{eq:parameterisation_gamma} of $\Gamma$ with $p=1$ (depicted in the left panels). The white regions correspond to $\sigma_{\text{max}} \leq 0$ and hence linearly stable regions. The black line separates linearly stable and unstable regions (and do not depend on the hyperdiffusion $\kappa_{4}$). The dotted grey line corresponds to the contour line $\Delta = 0$. The stars mark the parameter values of the cases studied in section~\ref{section:nonlinear_dynamics}. Note from the vertical axes that the lower panels correspond to the small $\text{Ri}_b$ region of the upper panels. \rev{Note also the difference between the scales of the two right panels. }}
    \label{fig:stability_plot_maximum_growth_rate}
\end{figure}

\begin{figure}
    \centerline{
    \includegraphics[width=\linewidth, trim={0 0 0 1cm}]{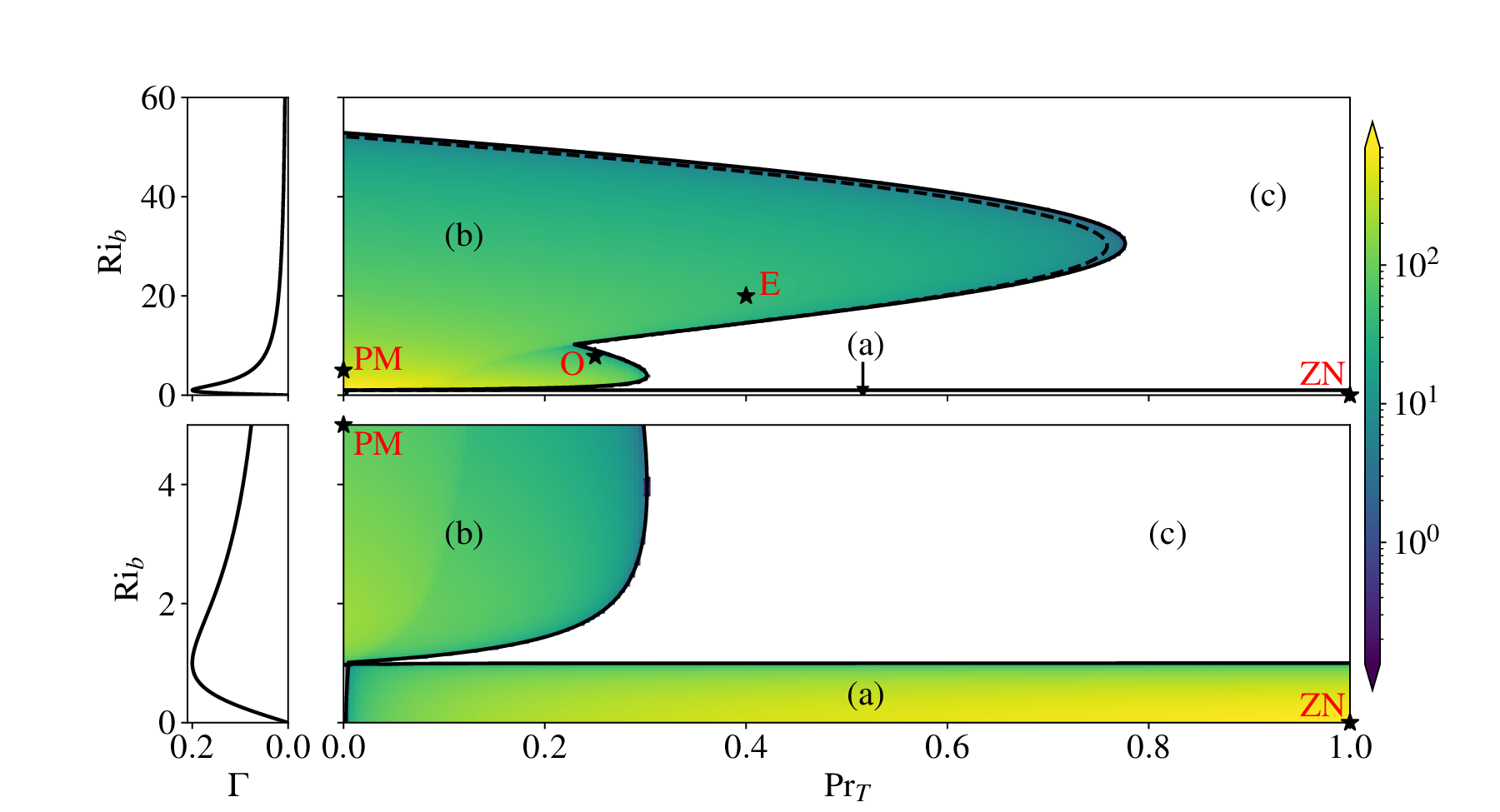}}
    \caption{Variation of the largest unstable wavenumber $k_{c}$ (on a logarithmic scale) with bulk Richardson number $\text{Ri}_{b}$ and turbulent Prandtl number $\text{Pr}_{T}$ for $\text{Pr} = 7$, $\text{Re} = 1000$, $\kappa_{4} = 10^{-7}$ and using parameterisation~\eqref{eq:parameterisation_gamma} of $\Gamma$ with $p=1$ (depicted in the left panels). The white regions correspond to $k_{c} = 0$ and hence linearly stable regions. The black line separates linearly stable and unstable regions. The dashed line corresponds to the $k_{c} = 2\pi$ contour line. Whereas parameters in the colored regions (regions (a) and (b)) are prone to \rev{staircase} formation, only parameters in region inside the dotted line will exhibit \rev{staircase formation} dynamics numerically. The stars correspond to the cases studied in section~\ref{section:nonlinear_dynamics}. Note from the vertical axes that the lower panels correspond to the small $\text{Ri}_b$ region of the upper panels. }
    \label{fig:stability_plot_maximum_unstable_wave_number}
\end{figure}

\rev{All in all, the boundaries between stable and unstable regions do not depend on the `strength' of the regularising hyperdiffusion (since $f$ and $C$ do not depend on $\kappa_{4}$). However, $\kappa_{4}$ does affect the magnitude of the growth rates. As a result, the system now has a largest unstable wavenumber (denoted $k_{c}$ in the subsequent and of order ${\cal O}(1/\text{Re}\kappa_{4})^{1/2}$ as shown previously) and a maximum growth rate $\sigma_{\text{max}}$ attained at a wavenumber that we will denote $k_{\text{max}}$. We plot these quantities for various values of the parameters in figures~\ref{fig:stability_plot_maximum_growth_rate} and~\ref{fig:stability_plot_maximum_unstable_wave_number}. The maximum growth rate $\sigma_{\text{max}}$ may be interpreted as a relevant time scale of \rev{staircase} formation whereas $k_{\text{max}}$ may be thought of as the length scale of the \rev{staircases} potentially forming, at least at early times, before subsequent coarsening through layer merger,  as we discuss further below. }

\rev{It is apparent from figures~\ref{fig:stability_plot_maximum_growth_rate} and~\ref{fig:stability_plot_maximum_unstable_wave_number} that the unstable region on the decreasing right flank of the $\Gamma$-curve (region (b)) divides into two distinct regions of relatively large $\sigma_{\text{max}}$ and $k_{c}$ separated by a gap of relatively small $\sigma_{\text{max}}$ and $k_{c}$, suggesting the existence of different types of unstable dynamics for $\text{Ri}_{b} \geq \text{Ri}_{b}^{m}$. A more precise description of these two dynamics can be given by considering in more detail the dispersion relation of the corrected system. More precisely, it can be written as: }
\begin{equation}
    \omega(k) = \frac{1}{2}\left[-2\text{i}\kappa_{4}k^{4} + \text{i}k^{2}f \pm \Delta_{h}(k)^{1/2}\right], 
\end{equation}
where $\Delta_{h}(k) := (-\text{i}\beta_{h})^{2} - 4\alpha_{h}\gamma_{h} = (-f^{2} - 4C)k^{4} = \Delta_{0}(k)$ ($\Delta_{0}$ has been defined in equation~\eqref{eq:dispersion_relation_no_hyperdiffusion}). Therefore, if $\Delta_{0}(k) > 0$, \rev{$\omega(k)$ has both an imaginary and a real part} and the component of vertical wavenumber $k$ of the solution of the linearised problem will hence be exponentially increasing or decreasing (depending on the sign of the imaginary part of $\omega(k)$) while oscillating with a frequency $\frac{1}{2}\Delta_{0}(k)^{1/2}$. \rev{On the contrary, if $\Delta_{0}(k) \leq 0$, then $\omega(k)$ is purely imaginary and the dynamic of the linearised solution associated with the wavenumber $k$ will be purely exponentially increasing or decreasing. Note that} $\Delta_{0}$ is of the sign of $\Delta := -f^{2} - 4C$, which depends only on the parameters $\text{Ri}_{b}$, $\text{Pr}_{T}$, $\text{Pr}$ and $\text{Re}$,  but crucially not on $k$ \rev{nor on $\kappa_{4}$}. We therefore expect different dynamics depending on the sign of $\Delta$: an `oscillatory' behaviour for parameters satisfying $\Delta > 0$ and a purely damped or exponentially growing one for $\Delta \leq 0$. \rev{(Note that these conditions on $\Delta$ correspond to the condition for the diffusion matrix of our linearised system (see section~\ref{section:link_with_diffusion}) to have or not eigenvalues with non-vanishing imaginary parts.)} Importantly, this result is independent of the addition of an hyperdiffusion correction. The contour line corresponding to $\Delta = 0$ is plotted on figure~\ref{fig:stability_plot_maximum_growth_rate}. Interestingly, it aligns with the gap of small $\sigma_{\text{max}}$ and $k_{c}$ mentioned above and shown in figures~\ref{fig:stability_plot_maximum_growth_rate} and~\ref{fig:stability_plot_maximum_unstable_wave_number}. This supports again the fact that at least two different types of unstable dynamics coexist in the unstable region (b). 

Using the above results, we can determine a relevant value for $\kappa_{4}$. More precisely, it is chosen so that the largest unstable wavenumber $k_{c}$ is of order or smaller than, in dimensionless form, $L^{\ast} / L_{K}^{\ast}$ where $L_{K}^{\ast} := ({\nu^{\ast}}^{3}/\Ast{\epsilon})^{1/4}$ is the Kolmogorov length scale.
\cpc{We choose this scale as it is the scale below which viscosity finally dissipates kinetic energy. Since the flows we are interested in typically have $\text{Pr} \gtrsim 1$, $L_K^\ast > L_B^\ast := L_K^{\ast}/\sqrt{\text{Pr}}$, where $L_{B}$ is the Batchelor scale at which fine structure in the scalar field is smoothed out by diffusivity. Therefore $L_K$ is a natural conservative scale to choose to regularise the build-up of perturbations at small scales.} We have shown that the largest unstable wavenumber is of order $\mathcal{O}(1/\text{Re}\kappa_{4})^{1/2}$ \rev{and, using the inertial scaling, $L^{\ast} / L_{K}^{\ast}$ is of order $\mathcal{O}(\text{Re}^{3/4})$. }Therefore, for $\text{Re} = \mathcal{O}(1000)$, we want $\kappa_{4} \gtrsim 10^{-8}$. For the purpose of our numerical experiment (section~\ref{section:nonlinear_dynamics}) and in order to form \rev{staircases} that are not too small nor too large, we \cpc{henceforth choose the conservative values} $\kappa_{4} = 10^{-5}$ or $10^{-7}$, depending on the particular choice of the parameters, as discussed further below. 

\section{Nonlinear dynamics}
\label{section:nonlinear_dynamics}

In this section, we numerically solve the \rev{regularised} dimensionless system~\eqref{eq:system_adim_hyperdiffusion} and compare the nonlinear dynamics to the \cpc{linear} stability analysis presented above. 

In order to solve~\eqref{eq:system_adim_hyperdiffusion}, boundary conditions need to be specified. In the following, we consider periodic boundary conditions for the shear $S$ and stratification $N^{2}$: 
\begin{equation}
    \forall t \geq 0, \; S(t, z=0) = S(t, z=1), \; N^{2}(t, z=0) = N^{2}(t, z=1). 
\end{equation}
These conditions \cpc{quantize} the range of admissible vertical wavenumbers $k$, \cpc{which are} now of the form $k = 2\pi n$ where $n = 0,\,1,\,\ldots$. \cpc{(in practice, $n$ will in fact be bounded above by $1/\mathrm{d}z$ where $\mathrm{d}z$ is the spatial grid size of our numerical calculations)}. Since the case $n=0$ \cpc{has zero} growth rate, if the largest unstable wavenumber $k_{c}$ (which exists thanks to the addition of hyperdiffusion) is smaller than $2\pi$, the system will be `numerically' linearly stable, \cpc{although it could of course} have been linearly unstable provided other boundary conditions were chosen. We plot the largest unstable wavenumber  for various values of the parameters in figures~\ref{fig:stability_plot_maximum_unstable_wave_number} as well as the contour line $k_{c} = 2\pi$. 

\cpc{Inspired by} the linear stability analysis (section~\ref{section:linear_stability_analysis}) we consider the following dimensionless initial conditions for three different choices with non-zero $\text{Ri}_b$ as marked on figure \ref{fig:stability_plot_maximum_unstable_wave_number} (cases PM, O and E, i.e. `Phillips Mechanism', `Oscillatory' and `Exponential'): 
\begin{equation}
    \forall z \in [0,1], \; S(t=0, z) = 1 + \Tilde{n}(z), \; N^{2}(t=0, z) = 1 + \Tilde{n}(z), 
\end{equation}
where $\Tilde{n}$ is a small random noise, normally distributed with $0$ mean and $0.01$ standard deviation. For the zero $\text{Ri}_b$ case (case ZN, i.e. `zero $N$'), ${N_{c}^{\ast}}^{2} = 0$ and we then set $N^{2}(t=0, z) = \text{max}(0, \Tilde{n}(z)) \geq 0$ so that the profile is always statically stably stratified. 

Using the above boundary and initial conditions,
system~\eqref{eq:system_adim_hyperdiffusion} can be solved using the method presented in appendix~\ref{appendix:numerical_scheme}. In order to \cpc{obtain} the velocity and the buoyancy fields from the computed shear and the stratification profiles, $S$ and $N^{2}$ are integrated over space. The integration constants are chosen so that the conservation of mass and momentum is respected: 
\begin{equation}
    \forall t \geq 0, \int_{0}^{1}b(t,z)\mathrm{d}z = \int_{0}^{1}b(0,z)\mathrm{d}z, \; \int_{0}^{1}u(t,z)\mathrm{d}z = \int_{0}^{1}u(0,z)\mathrm{d}z.
\end{equation}

\begin{figure}
    \centerline{
    \includegraphics[width=\linewidth]{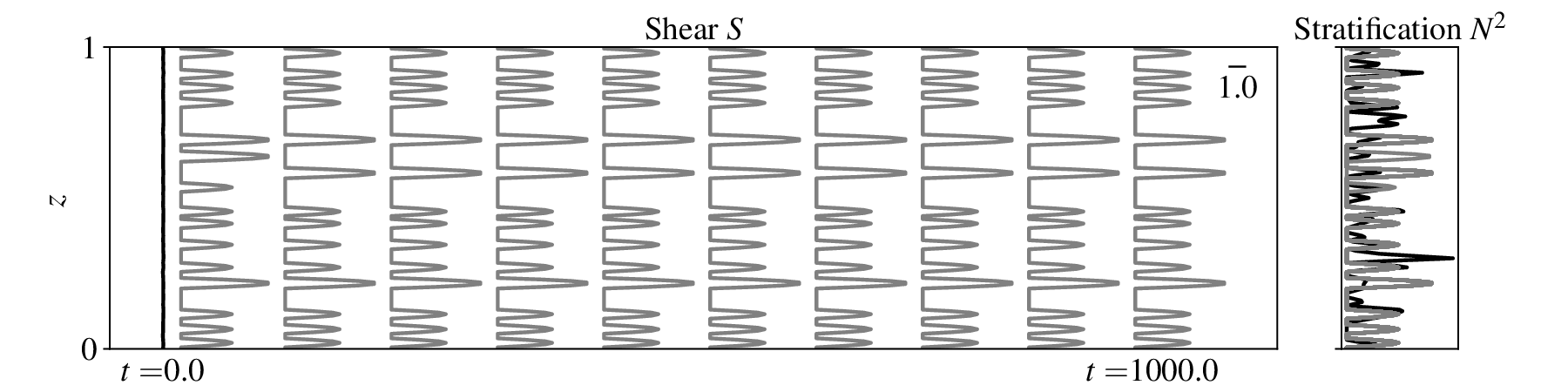}}
    \caption{Evolution of the shear and stratification for case N0 with $N_{c}^{2}=0$, $\text{Pr}_{T}=1$, $\text{Pr}=7$ and $\text{Re}=1000$. On the left panel, the horizontal axis is time and each profile is separated by $100$ dimensionless time units. Peaks in these profiles correpond to interfaces of enhanced vertical gradients separating well-mixed layers. The initial shear profile is depicted in black. \rev{The horizontal scale on the top right corner corresponds to a representative scale of variations of the shear $S$.} On the right panel, the initial condition (black line) as well as the profile at a later time (grey lines) \cpc{can be plotted} on the same axis, \cpc{demonstrating} that the \cpc{magnitude of the} perturbations on the buoyancy profile do not grow \cpc{significantly} above the initial perturbation.  (Note that in order to form \rev{staircases} that are big enough to be visible, we use the \cpc{higher} value of  $\kappa_{4} = 10^{-5}$.)}
    \label{fig:time_line_Pr_T_1_Rib_0}
\end{figure}

\begin{figure}
    \centerline{
    \includegraphics[width=\linewidth]{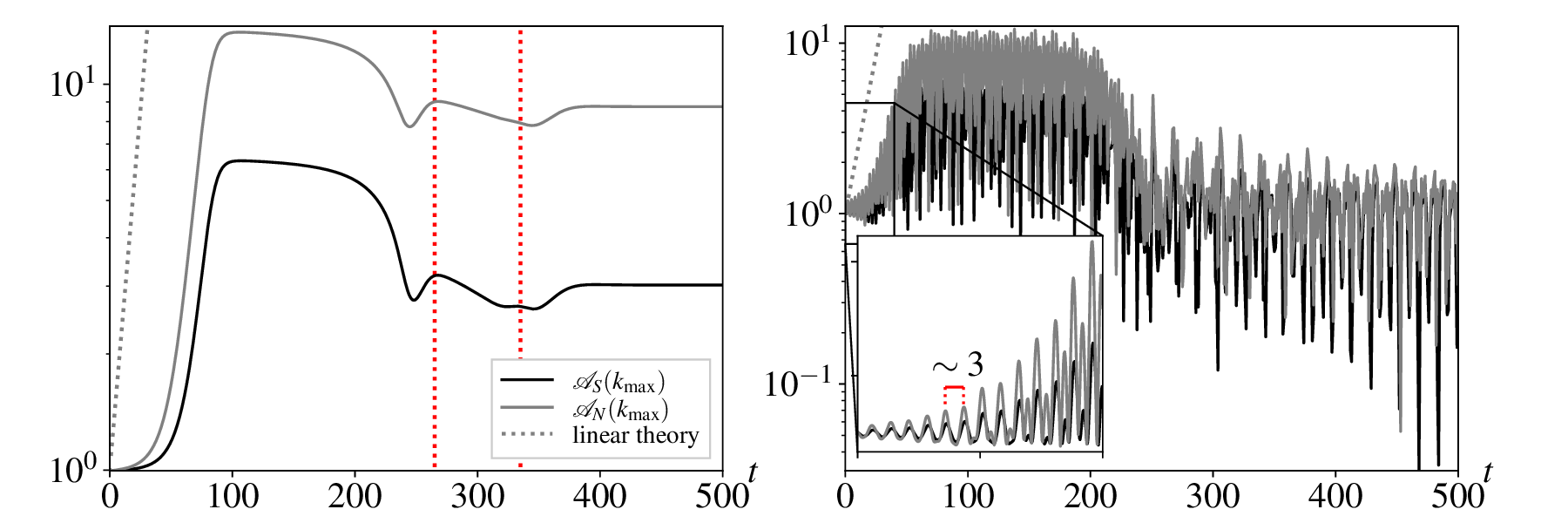}}
    \caption{Time evolution of the amplitude of the fastest growing mode (normalized by the initial amplitude) for: case E with $\text{Ri}_{b} = 20$, $\text{Pr}_{T} = 0.4$, $\text{Pr} = 7$, $\text{Re} = 1000$ and $\kappa_{4} = 10^{-7}$ (left panel); and case O with $\text{Ri}_{b} = 7.8$, $\text{Pr}_{T} = 0.25$, $\text{Pr} = 7$, $\text{Re} = 1000$ and $\kappa_{4} = 10^{-7}$ (right panel) for the shear (black line) and stratification (grey line). The dotted grey line corresponds to the evolution predicted by the linear theory. The red dotted lines correspond to the end of merging events. }
    \label{fig:time_evolution_fastest_growing_mode}
\end{figure}

\begin{figure}
    \centerline{
    \includegraphics[width=1.05\linewidth, trim={0 0 0 -1cm}]{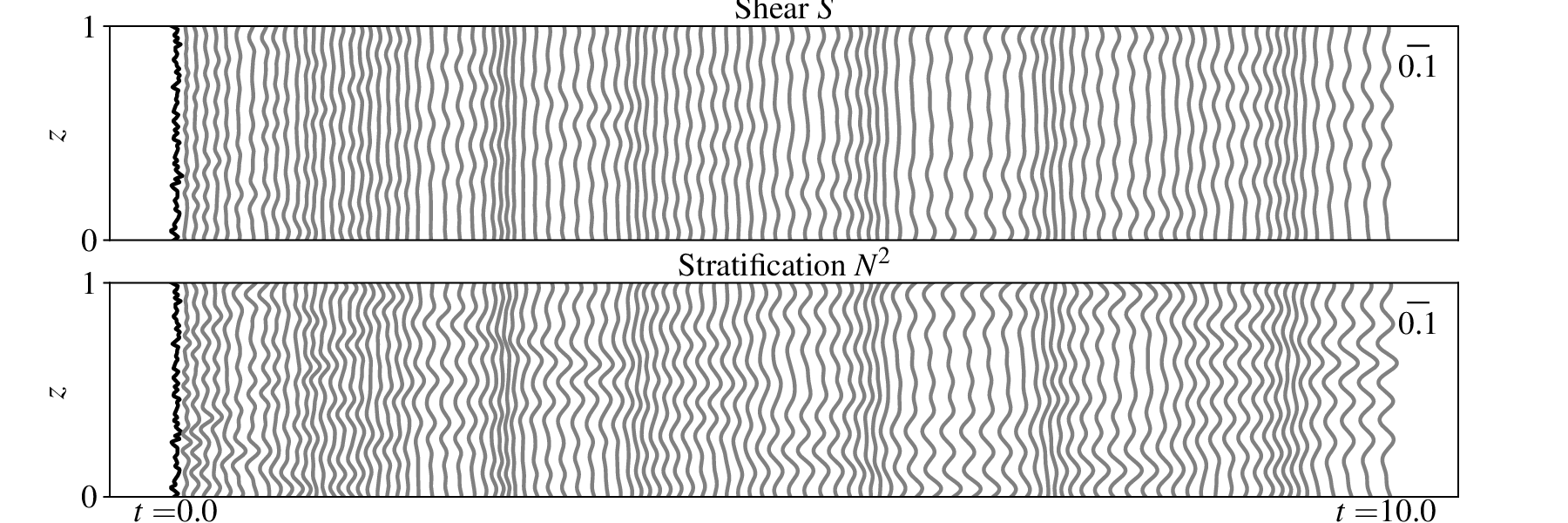}}
    \caption{Evolution of shear and stratification profiles for case O with $\text{Ri}_{b}=7.8$, $\text{Pr}_{T}=0.25$, $\text{Pr}=7$ and $\text{Re}=1000$. The black lines correspond to the initially disturbed profiles. On the top panel, the horizontal axis is time and each profile is separated by $0.1$ dimensionless time unit.  We use the \cpc{lower} value of  $\kappa_{4} = 10^{-7}$.}
    \label{fig:evolution_Pr_T_025_Rib_78_early_times}

    \centerline{
    \includegraphics[width=1.05\linewidth, trim={0 0 0 -1cm}]{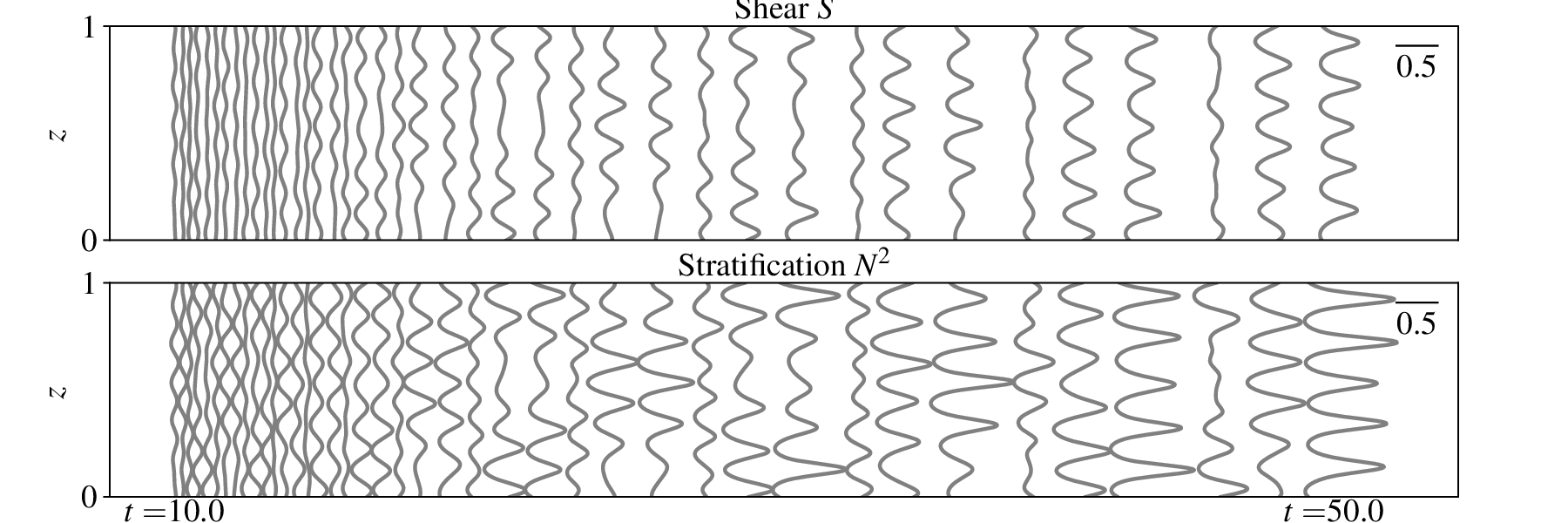}}
    \caption{\cpc{Later time evolution  in the same format as   figure~\ref{fig:evolution_Pr_T_025_Rib_78_early_times} (case O). The horizontal axis is again time and each profile is separated by $1$ dimensionless time unit. }}
    \label{fig:evolution_Pr_T_025_Rib_78_mid_times}

    \centerline{
    \includegraphics[width=1.05\linewidth, trim={0 0 0 -1cm}]{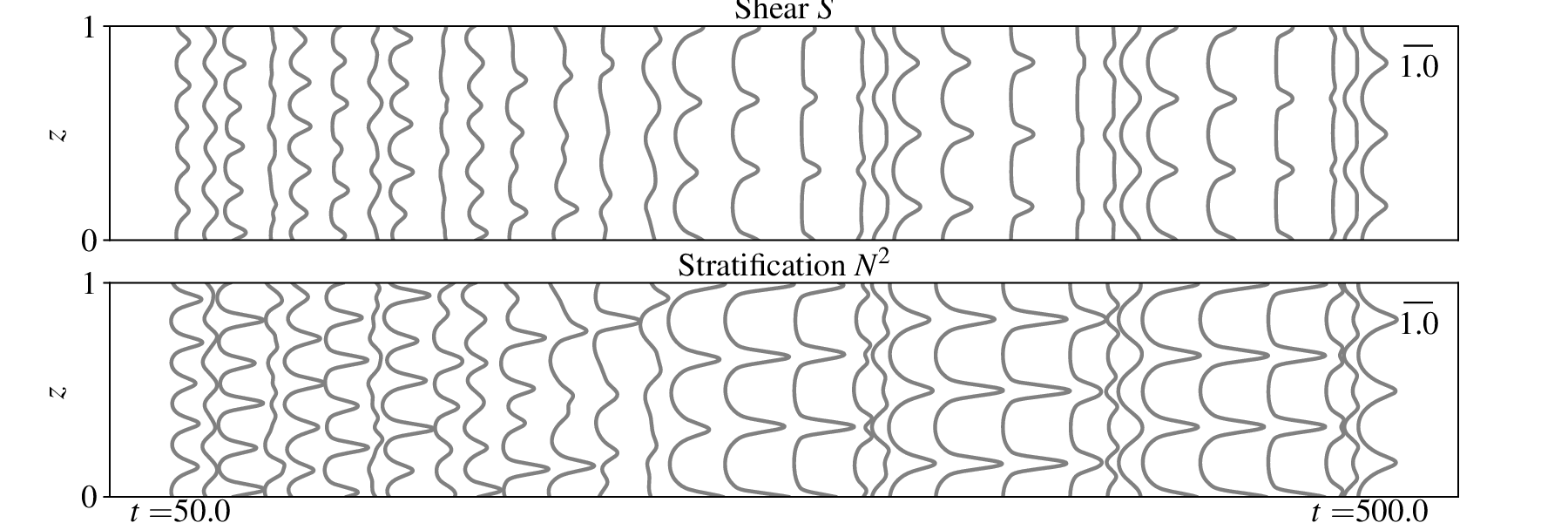}}
    \caption{Further later time evolution in the same format as  figure~\ref{fig:evolution_Pr_T_025_Rib_78_early_times} (case O). The horizontal axis is again time and each profile is separated by $15$ dimensionless time unit. }
    \label{fig:evolution_Pr_T_025_Rib_78_late_times}
\end{figure}

\begin{figure}
    \centering
    \begin{subfigure}[b]{\textwidth}
        \includegraphics[width=\linewidth]{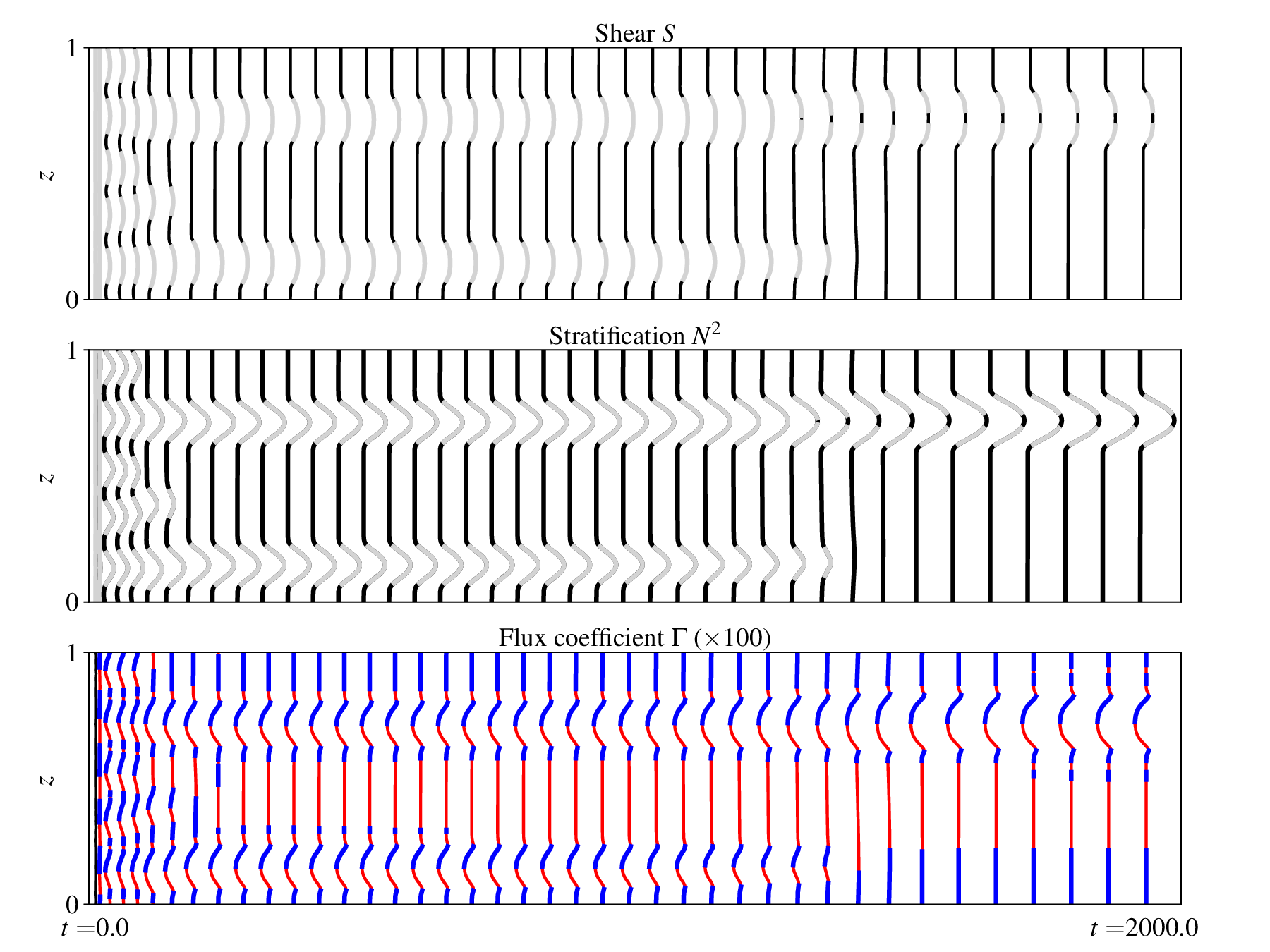}
    \end{subfigure}\\
    \begin{subfigure}[b]{\textwidth}
        \includegraphics[width=\linewidth]{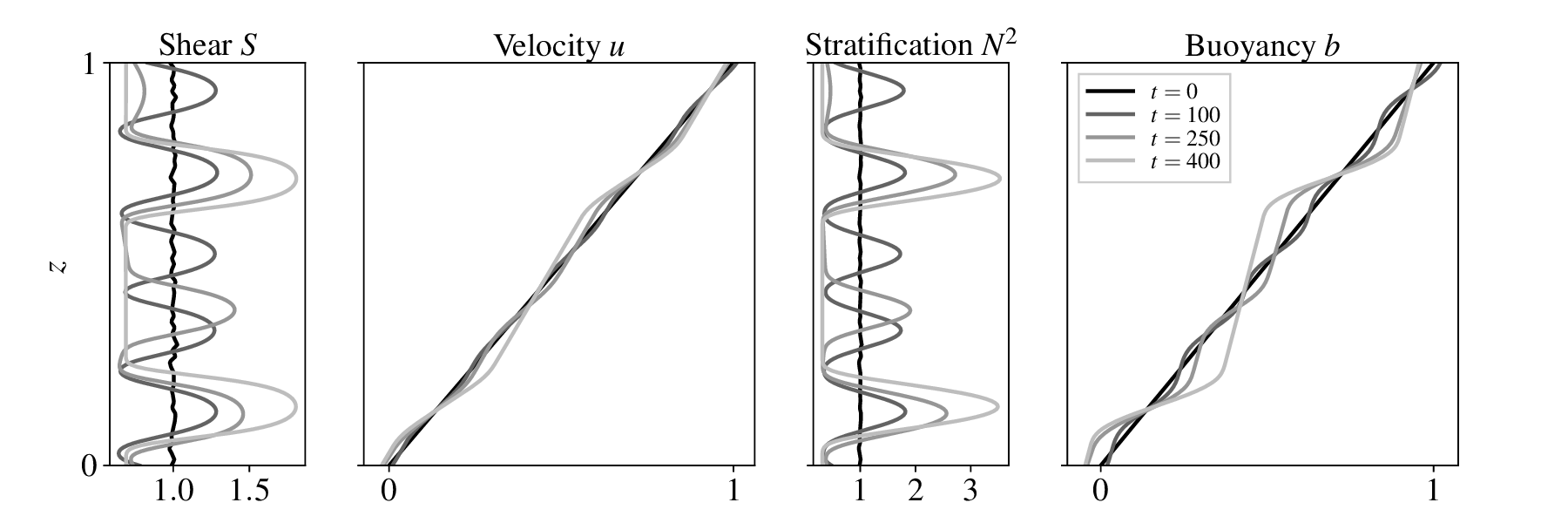}
    \end{subfigure}
    \caption{Evolution of the buoyancy, velocity and flux coefficient profiles for case E with $\text{Ri}_{b}=20$, $\text{Pr}_{T}=0.4$, $\text{Pr}=7$ and $\text{Re}=1000$. On the top three panels, the horizontal axis is time and each profile is separated by $50$ dimensionless time unit. On the shear and stratification profiles, the grey regions correspond to regions where the effective eddy diffusivity defined in appendix~\ref{appendix:non_linear_diffusion} is negative. On the flux coefficient profiles, red corresponds to upwelling while blue corresponds to downwelling. }
    \label{fig:evolution_Pr_T_04_Rib_20_late_times}
\end{figure}

We focus our attention on four sets of linearly unstable parameters, \rev{each with $\text{Re}=1000$ and $\text{Pr}=7$}, as marked with stars on figures~\ref{fig:stability_plot_maximum_growth_rate} and~\ref{fig:stability_plot_maximum_unstable_wave_number}: 
\begin{itemize}
    \item \textbf{Case ZN}: ${N_{c}^{\ast}}^{2} = 0$ and $\text{Pr}_{T} = 1$. This choice of parameters illustrates the limiting unstratified case ${N_{c}^{\ast}}^{2} = 0$ presented in section~\ref{section:N2_0}. We show the numerical results in figure~\ref{fig:time_line_Pr_T_1_Rib_0}. Here the perturbations in the buoyancy profile do not grow above their initial magnitude and layering occurs in the velocity profile.
    
    \item \textbf{Case PM}: $\text{Ri}_{b}=5$ (\cpc{on the} decreasing \cpc{right} flank of \cpc{the $\Gamma$-curve}) and $\text{Pr}_{T}=0$. This set of parameters illustrates the theoretical results presented in section~\ref{section:Pr_T_0} for $\text{Pr}_{T} = 0$. \cpc{We show} the numerical solution  in figure~\ref{fig:evolution_Pr_T_O_Rib_5}. Perturbations in the velocity profile are damped and the linear velocity profile (constant shear profile) is retrieved whereas perturbations in the buoyancy profile grow and form layers that eventually merge. 
    
\begin{figure}
    \centering
    \begin{subfigure}[b]{\textwidth}
        \includegraphics[width=\linewidth]{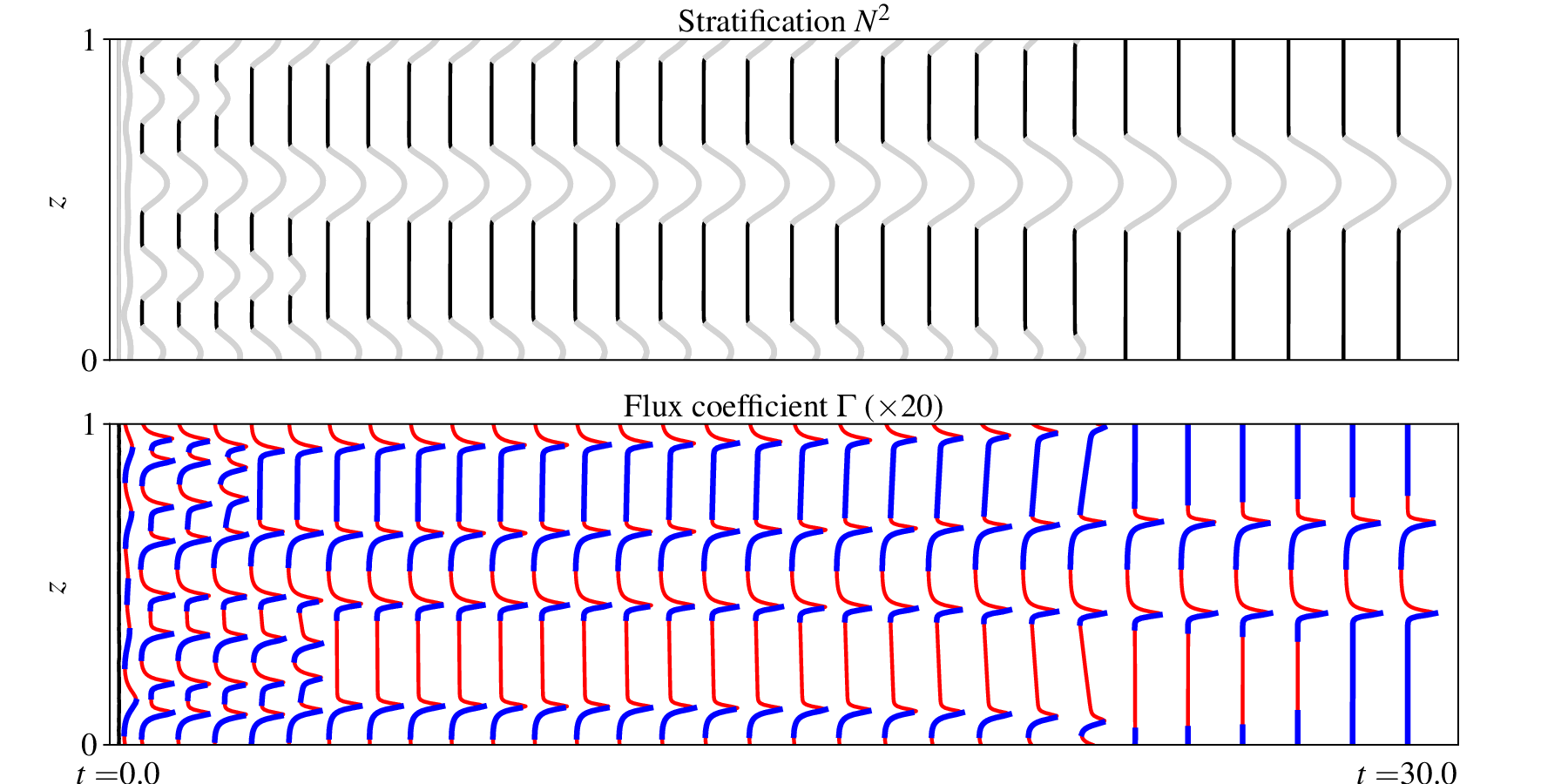}
    \end{subfigure}\\
    \begin{subfigure}[b]{\textwidth}
        \includegraphics[width=\linewidth]{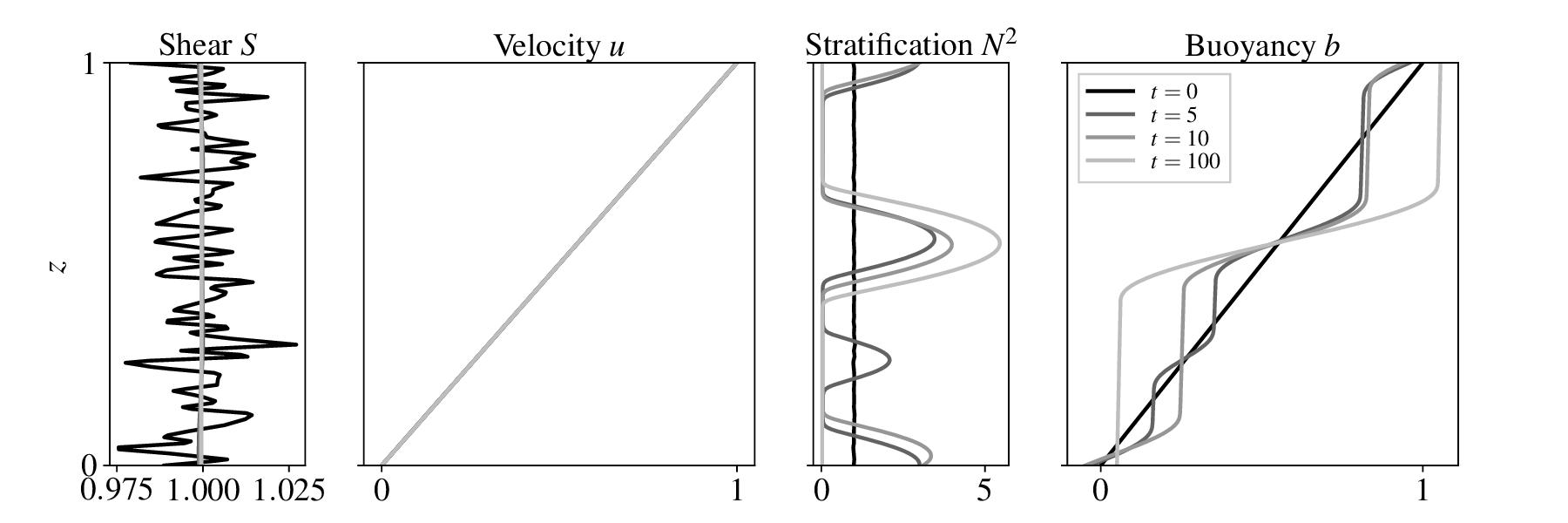}
    \end{subfigure}
    \caption{Evolution of the buoyancy, velocity and flux coefficient profiles for case PM with $\text{Ri}_{b}=5$, $\text{Pr}_{T}=0$, $\text{Pr}=7$ and $\text{Re}=1000$. On the bottom panels, the black lines correspond to the initially disturbed profiles (the same perturbation is used for both the velocity and buoyancy). On the top two panels, the horizontal axis is time and each profile is separated by one dimensionless time unit. On the $N^{2}$-profiles the grey regions correspond to regions where the effective eddy diffusivity defined in appendix~\ref{appendix:non_linear_diffusion} is negative. For the flux coefficient profiles, red corresponds to upwelling while blue corresponds to downwelling, unveiling the convergence/divergence of buoyancy flux patterns underlying the Phillips mechanism. (Note that in order to form \rev{staircases} that are big enough to be visible, we use the \cpc{higher} value of  $\kappa_{4} = 10^{-5}$.) 
    }
    \label{fig:evolution_Pr_T_O_Rib_5}
\end{figure}
    
    \item \textbf{Case O}: $\text{Ri}_{b}=7.8$ (decreasing right flank of the $\Gamma$-curve) and $\text{Pr}_{T}=0.25$. This set of parameters has been chosen in the unstable region (b) as shown in figure~\ref{fig:stability_plot_maximum_unstable_wave_number}, so that $\Delta > 0$. The maximum growth rate is $\sigma_{\text{max}} \simeq 0.09$ and attained for a wavenumber $k_{\text{max}} \simeq 30.4$. Therefore we  expect the development of structures of length-scale $\sim 0.2$ dimensionless space units in around $10$ dimensionless time units. Since $\Delta > 0$, we also expect some kind of oscillatory behaviour in the time evolution of the buoyancy and velocity profiles. We show the time evolution of the amplitude of the fastest growing mode (corresponding to $k_{\text{max}}$) as well as numerical profiles in  figures~\ref{fig:time_evolution_fastest_growing_mode},~\ref{fig:evolution_Pr_T_025_Rib_78_early_times},~\ref{fig:evolution_Pr_T_025_Rib_78_mid_times} and~\ref{fig:evolution_Pr_T_025_Rib_78_late_times}. After a transient phase, yet before the saturation of the instability, the perturbation appears to grow at the predicted rate simultaneously and concomitantly in both the shear and stratification profiles. Interestingly, \rev{staircases} seem to `pulse' with a period of approximately $3$ dimensionless time units, corresponding to the theoretical period $2\pi / (0.5\Delta_{0}(k_{\text{max}})^{1/2}) \simeq 3$ (see section~\ref{section:instability_properties}). Furthermore, the development of buoyancy and velocity staircases appears to be locked and in phase. The initial layers (before they start merging), have a length scale of $\sim 0.2$ dimensionless space units, \cpc{demonstrating} the \cpc{relevance} of the linear stability analysis. Similar dynamics are observed for other sets of linearly unstable parameters on the decreasing right flank of the $\Gamma$-curve satisfying $\Delta > 0$.
    
    \item \textbf{Case E}: $\text{Ri}_{b}=20$ \cpc{(also on the decreasing right flank of  the $\Gamma$-curve)} and $\text{Pr}_{T}=0.4$. This set of parameters lies on the linearly unstable region (b) and satisfies $\Delta \leq 0$. It has been chosen so that the unstable branch of the growth rate spectrum associated with this case is similar to the one associated with the previous case (see figure~\ref{fig:growth_rate_wave_number_dependence}). Therefore, the relevant time and length scales associated with the development of potential instabilities will be similar in both cases and we expect a structure of length scale $\sim 0.2$ dimensionless space units to appear. We show the time evolution of the amplitude of the fastest growing mode as well as numerical profiles in figures~\ref{fig:time_evolution_fastest_growing_mode} and~\ref{fig:evolution_Pr_T_04_Rib_20_late_times}. After a short transient (and before saturation of the instability), the fastest growing mode grows at the expected theoretical rate. The initial layers (before they start merging) have a length scale of $\sim 0.2$ dimensionless space units, once again as predicted by the linear theory. As the instability saturates, layers start to merge. No oscillations in time are observed, in line with $\Delta \leq 0$. Interestingly, and unlike the previous case where the perturbations in the stratification and the shear seemed to evolve concomitantly and \rev{staircases} \cpc{form} almost simultaneously and in phase in the buoyancy and velocity profiles, buoyancy \rev{staircases} seem to form slightly before velocity ones. Similar dynamics are observed for other sets of parameters on the decreasing right flank of the $\Gamma$-curve satisfying $\Delta \leq 0$. 
    \cpc{This behaviour is also} reminiscent of the case $\text{Pr}_{T} = 0$ \cpc{exhibiting the} Phillips mechanism and associated with the condition $C \geq 0$ (that implies $\Delta \leq 0$) where \rev{staircases} form exclusively in the buoyancy field.  
\end{itemize}

All in all, the unstable parameters region should be thought of as being divided into three subregions: a low $\text{Ri}_{b}$ region (corresponding to $\text{Ri}_{b} \ll 1$) where the dynamics is mostly shear-driven and where \rev{staircase} formation happens in the velocity profile since there are no buoyancy gradients to mix; an intermediate $\text{Ri}_{b}$ and small $\text{Pr}_{T}$ region (corresponding to $\text{Ri}_{b} \geq \text{Ri}_{b}^{m}$, $\Delta > 0$) where the dynamics is buoyancy- and shear-driven and where \rev{staircases form} almost simultaneously in both the buoyancy and velocity fields with \rev{staircase} `pulsation'; and an intermediate to large $\text{Ri}_{b}$ and small $\text{Pr}_{T}$ region (corresponding to $\text{Ri}_{b} \geq \text{Ri}_{b}^{m}$ and $\Delta \leq 0$) where the dynamics is again shear- and buoyancy-driven and \rev{staircases develop without `pulsation' before merging as the instability saturates. }

The nonlinear dynamics also pinpoint the qualitatively different mixing happening in the well-mixed layers and in the strongly stratified interfaces separating layers. Inside the layers, the density anomalies are smoothed and mixing can therefore be described by an appropriately defined positive eddy diffusivity (see appendix~\ref{appendix:non_linear_diffusion}). In the interfaces, such an eddy diffusivity becomes \cpc{formally} negative (see figure~\ref{fig:evolution_Pr_T_O_Rib_5} and~\ref{fig:evolution_Pr_T_04_Rib_20_late_times} for instance) and the mixing  is therefore in some sense `antidiffusive', in the specific sense that it appears to sharpen the buoyancy gradients by scouring the interface, as suggested by the presence of local maxima of $\Gamma$ at the borders of density interfaces (although further analysis and direct numerical simulations are undoubtedly needed to confirm this point). Similarly, the observation that inside the interfaces the flux coefficient is minimal supports the hypothesis that density staircases are barriers to mixing. 

\section{Discussion}
\label{section:discussion}

In this paper, we have derived a reduced order model aiming at describing the formation and evolution of density staircases in sheared and (stably) stratified turbulent flows. Following the ideas of \citet{phillips1972turbulence} and \citet{posmentier1977generation}, we have parameterised the turbulence using flux-gradient models. Using this framework, we have determined regions in the parameter space $(\text{Ri}_{b}, \text{Pr}_{T}, \text{Pr}, \text{Re})$ prone to \rev{staircase} formation. Crucially, these regions depend on the monotonicity of variation of the flux coefficient $\Gamma$ with the bulk Richardson number $\text{Ri}_{b}$. Since experimental, observational and numerical evidence seem to indicate that $\Gamma$ increases with $\text{Ri}_{b}$ up to some critical value $\text{Ri}_{b}^{m}$ and \cpc{plausibly} decreases for $\text{Ri}_{b} \geq \text{Ri}_{b}^{m}$ \citep{Linden_1979, Linden_1980, TheRelationshipbetweenFluxCoefficientandEntrainmentRatioinDensityCurrents}, the staircase `instability' depends on the size of $\text{Ri}_{b}$ compared to $\text{Ri}_{b}^{m}$. Most importantly, we have also presented theoretical evidence that \cpc{this instability} depends on the turbulent Prandtl number $\text{Pr}_{T}$. 

On the increasing \cpc{left} flank of the $\Gamma$-curve, the instability occurs for $\text{Pr}_{T}$ above a \cpc{(very small)} given threshold, found \cpc{to be} around $0.001$ for the case with $\Gamma(\text{Ri}_{b}) \propto \text{Ri}_{b}$, $\text{Pr}=7$ and $\text{Re} = \mathcal{O}(1000)$. Therefore, for \cpc{sufficiently small} $\text{Pr}_{T} \ll 1$, $\text{Ri}_{b} < \text{Ri}_{b}^{m}$ is stable to \rev{staircase} formation, retrieving Phillips result that stratification needs to be sufficiently large (i.e. on the decreasing right flank of the $\Gamma$-curve) to be prone to staircase formation. However, for larger \cpc{(though still small)} values of $\text{Pr}_{T}$, \rev{staircase} instabilities can \cpc{actually} be triggered in weakly stratified flows (in the sense $\text{Ri}_{b} < \text{Ri}_{b}^{m}$, i.e. on the increasing left flank of the $\Gamma$-curve) in the presence of buoyancy- and shear-driven turbulence. 

Conversely, on the decreasing right flank of the $\Gamma$-curve, the instability occurs for sufficiently small turbulent Prandtl numbers and moderate to large bulk Richardson numbers. More precisely, for relevant oceanic parameters, \rev{staircase formation} via the Phillips mechanism is only possible within this model for $\text{Pr}_{T} \lesssim 0.5 - 0.8$.  The existence of this upper bound on the turbulent Prandtl number, that importantly has been shown to depend strongly on $\text{Pr}$ and weakly on the precise parameterisation of the turbulent fluxes (in the sense that it depends only on the rate  of the decrease of $\Gamma$ with $\text{Ri}_{b}$) but not on $\text{Re}$ (for non-zero values of the molecular diffusivity $\Ast{\nu}$) \rev{nor on the scalings for the dissipation rate of turbulent kinetic energy discussed in this work}, confirms that the Phillips mechanism for \rev{staircase} formation in strongly stratified flows is only valid for small values of $\text{Pr}_{T}$ in the presence of both buoyancy- and shear-driven turbulence. It also suggests that \rev{staircase} formation is not favoured in ocean interiors, as empirically observed. Indeed, the turbulent Prandtl number in stably stratified turbulence is usually found to be  $\text{Pr}_T \gtrsim 0.7$ \citep{venayagamoorthy2010turbulent} and observational data (see figure~\ref{fig:intro}) supports the fact that ocean interiors are sufficiently strongly stratified, suggesting that these regions are not in a favourable regime of parameter for staircase formation via the Phillips mechanism. 


Considering further the decreasing right flank of the $\Gamma$-curve, as the molecular Prandtl number increases, the upper bound on $\text{Pr}_{T}$ increases, favouring \rev{staircase} formation as discussed in \citet{taylor_zhou_2017}. The upper bound on $\text{Pr}_{T}$ reaches values of order $\mathcal{O}(100)$ for $\text{Pr} = 700$ (salty water), consistently with the fact that \rev{staircase} formation has often been observed in laboratory experiments \rev{using salinity gradients rather than temperature gradients. This also suggests that staircase formation could be favoured in regions of the ocean where stratification is salt dominated, such as estuaries~\citep{holleman2016stratified}}. Finally, in the limit $\Ast{\nu} = 0$ (i.e. $\text{Re} \rightarrow \infty$), the upper bound on $\text{Pr}_{T}$ for instability is smaller than $1/2$, regardless of the form of $\Gamma$. This result, independent of the explicit form of $\Gamma$, supports again \rev{the fact that the Phillips mechanism for staircase formation in strongly stratified flows seems to survive only in the limit of small turbulent Prandtl numbers and that density staircase formation via this mechanism is not favoured in the presence of buoyancy- and shear-driven turbulence in relatively strongly stratified flows. } 


The nonlinear dynamics following the initial linear instability growth exhibit various interesting properties. For flows with unstable parameters on the decreasing right flank of the $\Gamma$-curve, the non-linear behaviour seems to be divided into two categories. For flows with parameters such that $\Delta > 0$ (see section~\ref{section:instability_properties}) a \rev{staircase} instability appears to develop simultaneously in both the buoyancy and velocity fields which forms layers that pulse and merge as time evolves. Conversely, for flows with parameters such that $\Delta \leq 0$ staircases develop without pulsing and merge as the instability saturates, reminiscent of the purely buoyancy-driven mechanism that occurs for $\text{Pr}_{T} = 0$, a case that is equivalent to the Phillips mechanism as formulated in~\citep{phillips1972turbulence} (section~\ref{section:Pr_T_0}) and for which the instability is also associated with the condition $\Delta \leq 0$. 

More generally, the nonlinear evolution of the layers underlines the qualitative differences between the mixing expected in the presence or absence of density staircases. In the absence of density staircases, the mixing is purely diffusive in the sense that it smooths density gradients and can be modeled by an appropriately defined positive eddy diffusivity (see appendix~\ref{appendix:non_linear_diffusion}). On the contrary, the interfaces between layers are characterised by a negative effective eddy diffusivity and the mixing process at hand scours density interfaces, sharpens density gradients and hence is in some sense `antidiffusive'. Since antidiffusive problems are both mathematically and numerically challenging, this raises intricate parameterisation issues for flux-gradient based models. 

There are of course several limitations of our model. Firstly, our model is relevant to regions of the ocean where double-diffusive effects (due to the presence of gradients of both temperature and salinity for instance) are negligible but breaks down in regions of the world's oceans where double-diffusion becomes prominent such as in polar regions or the Mediterranean Sea. Similarly, our model breaks down near boundaries where boundary effects might become significant. Secondly, it is important to remember that $\Gamma$ cannot be parameterised in terms of the Richardson number only. It depends on other parameters, such as the buoyancy Reynolds number $\text{Re}_{b} = \Ast{\epsilon} / (\nu^\ast \Ast{N}^{2})$ \citep{shih_koseff_ivey_ferziger_2005, salehipour2016new, mashayek2017efficiency}. \rev{Thirdly, we considered the different dimensionless control parameters of our system as free parameters that crucially were independent of each other and of the dynamical quantities. Several studies suggest however dependence of the turbulent Prandtl number $\text{Pr}_{T}$ with, for instance, the gradient or bulk Richardson number (see \citet{venayagamoorthy2010turbulent} or \citet{katul2014two} for more detailed discussions). Since our goal was to explore the full parameter space but not to assess whether it was indeed entirely accessible, we did not take these relationships into account. However, we expect that enforcing such constraints would restrict the range of accessible parameters but not change our linear stability results and hence would not alter the main conclusions of our work. } Similarly, our analysis has considered ranges of parameters prone to \rev{staircase} formation provided they can sustain turbulence (so that the considered scalings for $\Ast{\epsilon}$ hold). We however did not assess whether the full parameter space considered here could actually maintain turbulence, a study that is beyond the scope of this work. \rev{Similarly, our model assumes constant $\Ast{\epsilon}$ and focuses therefore on patches of relatively homogeneous in space and sustained in time turbulence.} However, the robustness of our results with regard to various scalings for $\Ast{\epsilon}$ (section~\ref{section:other_possible_scalings}) \rev{as well as to the size of $\Ast{\epsilon}$ itself} suggests that the model presented here is relevant to both shear-dominated and buoyancy-dominated turbulent regimes \citep{mater2014unifying} as well as to weakly and strongly stratified regimes \citep{garanaik2019inference}. 


\backsection[Acknowledgements]{For the purpose of open access, the authors have applied a Creative Commons Attribution (CC BY) licence to any Author Accepted Manuscript version arising from this submission. We kindly thank Lois Baker for giving us access to the data used to generate the first figure of this work. }

\backsection[Funding]{This project has received funding from the European Union’s Horizon 2020 research and innovation program under the Marie Skłodowska-Curie grant agreement N°956457. A.M. acknowledges support from National Environmental Research Council
(NE/P018319/1). }

\backsection[Declaration of interests]{The authors report no conflict of interest. 
}


\backsection[Author ORCID]{C. P. Caulfield, https://orcid.org/0000-0002-3170-9480, 
\\ A. Mashayek https://orcid.org/0000-0002-8202-3294. }


\appendix

\section{}\label{appendix:non_linear_diffusion}

Similarly to what we have done in section~\ref{section:link_with_diffusion}, we recast the full (nonlinear) problem~\eqref{eq:system_adim} as a diffusion problem. To do so, note that it can be put into the following matrix form: 
\begin{equation}
\renewcommand{\arraystretch}{1.3}
    \left[\begin{array}{c}
         \partial_{t}b  \\
         \partial_{t}u 
    \end{array}\right] 
    = \mathsfbi{D}_{nl}
    \left[\begin{array}{c}
         \partial_{zz}b  \\
         \partial_{zz}u 
    \end{array}\right],
\end{equation}
where: 
\begin{equation}
    \mathsfbi{D}_{nl} = \left[\begin{array}{cc}
         \frac{\Gamma'(\text{Ri}_{b}\text{Ri})}{(\partial_{z}u)^2} + \frac{1}{\text{Pr}\text{Re}} & -2\frac{\Gamma'(\text{Ri}_{b}\text{Ri})\text{Ri}}{\partial_{z}u} \\
         \frac{\text{Pr}_{T}}{\text{Ri}_{b}\text{Ri}^{2}(\partial_{z}u)^{3}}[\text{Ri}_{b}\text{Ri}\Gamma'(\text{Ri}_{b}\text{Ri}) - \Gamma(\text{Ri}_{b}\text{Ri})] & \frac{\text{Pr}_{T}}{\text{Ri}_{b}\text{Ri}\partial_{z}u}[-2\text{Ri}_{b}\text{Ri}\Gamma'(\text{Ri}_{b}\text{Ri}) + \Gamma(\text{Ri}_{b}\text{Ri})] + \frac{1}{\text{Re}}
    \end{array}\right]. 
\end{equation}
The matrix $\mathsfbi{D}_{\text{nl}}$ is the nonlinear diffusion matrix associated to our problem. The real part of the eigenvalues of this matrix can be interpreted as effective eddy diffusivities of the system. Since the sign of these real parts is related to the sign of the trace $\text{Tr}(\mathsfbi{D}_{\text{nl}})$ and determinant $\text{det}(\mathsfbi{D}_{\text{nl}})$ of $\mathsfbi{D}_{\text{nl}}$, regions where $\text{Tr}(\mathsfbi{D}_{\text{nl}}) < 0$ or $\text{det}(\mathsfbi{D}_{\text{nl}}) < 0$ will be prone to antiduffisive dynamics that will sharpen density interfaces. (Note that these quantities are defined locally in space.) Note that $-f$ and $-C$ are the zero-th order approximation of these quantities, linking the linear dynamics to the nonlinear dynamics.

\section{}\label{appendix:numerical_scheme}
Let us formally write the system of equations~\eqref{eq:system_adim_hyperdiffusion} in the following form: 
\begin{equation}
    \partial_{t}y = \partial_{zz}[f(y)] - \kappa_{4}\partial_{z}^{4}y.
\end{equation}
We first discretize the above in space using second order in space schemes and obtain: 
\begin{equation}
    \partial_{t}y_{i} = \frac{1}{\mathrm{d}z^{2}}[f(y_{i+1}) - 2f(y_{i}) + f(y_{i-1})] + \frac{1}{\mathrm{d}z^{4}}[y_{i+2} -4y_{i+1} + 6y_{i} - 4y_{i-1} + y_{i-2}], 
\end{equation}
where $i \in \{2,\cdots, N-2\}$ are the indices of the grid points, $\mathrm{d}z$ is the spacing between grid points and $\mathbf{y}(t) = (y_{0}(t),\cdots,y_{N}(t))^{\top}$ are the approximate values of $y(t)$ at the grid points. The formulae for $i \in \{0, 1, N-1, N\}$ depend on the boundary conditions used in $z=0$ and $z=1$. We have considered periodic boundary conditions in our analysis. The above can be put into a matrix form: 
\begin{equation}
    \partial_{t}\mathbf{y} = \mathsfbi{A}_{\mathrm{d}z}(\mathbf{y}), 
\end{equation}
with $\mathsfbi{A}_{\mathrm{d}z}: \mathds{R}^{N+1} \rightarrow \mathcal{M}^{N+1,N+1}(\mathds{R})$. This is a system of $N+1$ ODEs. We can now use an appropriate time-stepping scheme to solve the problem numerically. We have used the BDF method with adaptive step-size from the python library scipy in order to  resolve accurately the stiff dynamics that appear as \rev{staircases} form.  

As \rev{staircases} form, the shear $S$ might become close to zero. This can introduce inappropriate divisions by zero in the definition of the Richardson number and lead to numerical difficulties. To avoid this issue, we consider $\text{Ri} = \frac{N^{2}}{S^{2} + \eta}$ where $\eta$ is a small parameter. We use $\eta = 10^{-9}$ in our simulations.

\bibliographystyle{jfm}
\bibliography{main}

\begin{thebibliography}{39}
\expandafter\ifx\csname natexlab\endcsname\relax\def\natexlab#1{#1}\fi
\def\au#1{#1} \def\ed#1{#1} \def\yr#1{#1}\def\at#1{#1}\def\jt#1{\textit{#1}}
  \def\bt#1{#1}\def\bvol#1{\textbf{#1}} \def\vol#1{#1} \def\pg#1{#1}
  \def\publ#1{#1}\def\arxiv#1{#1}\def\org#1{#1}\def\st#1{\textit{#1}}

\bibitem[Balmforth {\em et~al.\/}(1998)Balmforth, Smith \&
  Young]{balmforth1998dynamics}
{\sc \au{Balmforth, NJ}, \au{Smith, Stefan G~Llewellyn} \& \au{Young, WR}}
  \yr{1998}  \at{Dynamics of interfaces and layers in a stratified turbulent
  fluid}.  \jt{Journal of Fluid Mechanics}  \bvol{355},  \pg{329--358}.

\bibitem[Barenblatt {\em et~al.\/}(1993)Barenblatt, Bertsch, Dal~Passo,
  Prostokishin \& Ughi]{barenblatt1993mathematical}
{\sc \au{Barenblatt, GI}, \au{Bertsch, M}, \au{Dal~Passo, R}, \au{Prostokishin,
  VM} \& \au{Ughi, Maura}} \yr{1993}  \at{A mathematical model of turbulent
  heat and mass transfer in stably stratified shear flow}.  \jt{Journal of
  Fluid Mechanics}  \bvol{253},  \pg{341--358}.

\bibitem[Billant \& Chomaz(2001)]{billant2001self}
{\sc \au{Billant, Paul} \& \au{Chomaz, Jean-Marc}} \yr{2001}
  \at{Self-similarity of strongly stratified inviscid flows}.  \jt{Physics of
  fluids}  \bvol{13}~(6),  \pg{1645--1651}.

\bibitem[Bouffard \& Boegman(2013)]{bouffard2013diapycnal}
{\sc \au{Bouffard, Damien} \& \au{Boegman, Leon}} \yr{2013}  \at{A diapycnal
  diffusivity model for stratified environmental flows}.  \jt{Dynamics of
  Atmospheres and Oceans}  \bvol{61},  \pg{14--34}.

\bibitem[Caulfield(2021)]{Annurev_Colm_Caulfield}
{\sc \au{Caulfield, C.P.}} \yr{2021}  \at{Layering, instabilities, and mixing
  in turbulent stratified flows}.  \jt{Annual Review of Fluid Mechanics}
  \bvol{53}~(1),  \pg{113--145},  \arxiv{arXiv:
  https://doi.org/10.1146/annurev-fluid-042320-100458}.

\bibitem[Cope {\em et~al.\/}(2020)Cope, Garaud \&
  Caulfield]{cope_garaud_caulfield_2020}
{\sc \au{Cope, Laura}, \au{Garaud, P.} \& \au{Caulfield, C.~P.}} \yr{2020}
  \at{The dynamics of stratified horizontal shear flows at low péclet number}.
   \jt{Journal of Fluid Mechanics}  \bvol{903},  \pg{A1}.

\bibitem[Garanaik \& Venayagamoorthy(2019)]{garanaik2019inference}
{\sc \au{Garanaik, Amrapalli} \& \au{Venayagamoorthy, Subhas~K}} \yr{2019}
  \at{On the inference of the state of turbulence and mixing efficiency in
  stably stratified flows}.  \jt{Journal of Fluid Mechanics}  \bvol{867},
  \pg{323--333}.

\bibitem[Garaud {\em et~al.\/}(2015)Garaud, Medrano, Brown, Mankovich \&
  Moore]{garaud2015excitation}
{\sc \au{Garaud, P}, \au{Medrano, M}, \au{Brown, JM}, \au{Mankovich, C} \&
  \au{Moore, K}} \yr{2015}  \at{Excitation of gravity waves by fingering
  convection, and the formation of compositional staircases in stellar
  interiors}.  \jt{The Astrophysical Journal}  \bvol{808}~(1),  \pg{89}.

\bibitem[Holleman {\em et~al.\/}(2016)Holleman, Geyer \&
  Ralston]{holleman2016stratified}
{\sc \au{Holleman, RC}, \au{Geyer, WR} \& \au{Ralston, DK}} \yr{2016}
  \at{Stratified turbulence and mixing efficiency in a salt wedge estuary}.
  \jt{Journal of Physical Oceanography}  \bvol{46}~(6),  \pg{1769--1783}.

\bibitem[Ivey {\em et~al.\/}(1998)Ivey, Imberger \& Koseff]{Ivey_1998}
{\sc \au{Ivey, Gregory}, \au{Imberger, Jorg} \& \au{Koseff, J.R.}} \yr{1998}
  {\em Buoyancy fluxes in a stratified fluid\/},  \st{Coastal and Estuarine
  Studies},  \vol{vol.~54},  \pg{pp. 377--388}.  \publ{United States: American
  Geophysical Union}.

\bibitem[Ivey \&
  Imberger(1991)]{OntheNatureofTurbulenceinaStratifiedFluidPartITheEnergeticsofMixing}
{\sc \au{Ivey, G.~N.} \& \au{Imberger, J.}} \yr{1991}  \at{On the nature of
  turbulence in a stratified fluid. part i: The energetics of mixing}.
  \jt{Journal of Physical Oceanography}  \bvol{21}~(5),  \pg{650 -- 658}.

\bibitem[Katul {\em et~al.\/}(2014)Katul, Porporato, Shah \&
  Bou-Zeid]{katul2014two}
{\sc \au{Katul, Gabriel~G}, \au{Porporato, Amilcare}, \au{Shah, Stimit} \&
  \au{Bou-Zeid, Elie}} \yr{2014}  \at{Two phenomenological constants explain
  similarity laws in stably stratified turbulence}.  \jt{Physical Review E}
  \bvol{89}~(2),  \pg{023007}.

\bibitem[{Kay} \& {Jay}(2003)]{Kay_&_Jay_2003}
{\sc \au{{Kay}, David~J.} \& \au{{Jay}, David~A.}} \yr{2003}  \at{{Interfacial
  mixing in a highly stratified estuary 1. Characteristics of mixing}}.
  \jt{Journal of Geophysical Research (Oceans)}  \bvol{108}~(C3),  \pg{3072}.

\bibitem[Kranenburg(1980)]{kranenburg1980stability}
{\sc \au{Kranenburg, C}} \yr{1980}  \at{On the stability of turbulent
  density-stratified shear flow}.  \jt{Journal of Physical Oceanography}
  \bvol{10}~(7),  \pg{1131--1133}.

\bibitem[Linden(1979)]{Linden_1979}
{\sc \au{Linden, P.~F.}} \yr{1979}  \at{Mixing in stratified fluids}.
  \jt{Geophysical \& Astrophysical Fluid Dynamics}  \bvol{13}~(1),  \pg{3--23},
   \arxiv{arXiv: https://doi.org/10.1080/03091927908243758}.

\bibitem[Linden(1980)]{Linden_1980}
{\sc \au{Linden, P.~F.}} \yr{1980}  \at{Mixing across a density interface
  produced by grid turbulence}.  \jt{Journal of Fluid Mechanics}
  \bvol{100}~(4),  \pg{691–703}.

\bibitem[Ma \& Peltier(2021)]{Ma_Peltier_2021_Gamma_instability}
{\sc \au{Ma, Yuchen} \& \au{Peltier, W.~R.}} \yr{2021}  \at{Gamma instability
  in an inhomogeneous environment and salt-fingering staircase trapping:
  Determining the step size}.  \jt{Phys. Rev. Fluids}  \bvol{6},  \pg{033903}.

\bibitem[Mashayek {\em et~al.\/}(2022)Mashayek, Baker, Cael \&
  Caulfield]{mashayek2022marginal}
{\sc \au{Mashayek, A}, \au{Baker, LE}, \au{Cael, BB} \& \au{Caulfield, CP}}
  \yr{2022}  \at{A marginal stability paradigm for shear-induced diapycnal
  turbulent mixing in the ocean}.  \jt{Geophysical Research Letters}
  \bvol{49}~(2),  \pg{e2021GL095715}.

\bibitem[Mashayek {\em et~al.\/}(2017)Mashayek, Salehipour, Bouffard,
  Caulfield, Ferrari, Nikurashin, Peltier \& Smyth]{mashayek2017efficiency}
{\sc \au{Mashayek, A}, \au{Salehipour, H}, \au{Bouffard, D}, \au{Caulfield,
  CP}, \au{Ferrari, R}, \au{Nikurashin, M}, \au{Peltier, WR} \& \au{Smyth, WD}}
  \yr{2017}  \at{Efficiency of turbulent mixing in the abyssal ocean
  circulation}.  \jt{Geophysical Research Letters}  \bvol{44}~(12),
  \pg{6296--6306}.

\bibitem[Mater \& Venayagamoorthy(2014)]{mater2014unifying}
{\sc \au{Mater, Benjamin~D} \& \au{Venayagamoorthy, Subhas~Karan}} \yr{2014}
  \at{A unifying framework for parameterizing stably stratified shear-flow
  turbulence}.  \jt{Physics of Fluids}  \bvol{26}~(3),  \pg{036601}.

\bibitem[Oglethorpe {\em et~al.\/}(2013)Oglethorpe, Caulfield \&
  Woods]{oglethorpe2013spontaneous}
{\sc \au{Oglethorpe, RLF}, \au{Caulfield, CP} \& \au{Woods, Andrew~W}}
  \yr{2013}  \at{Spontaneous layering in stratified turbulent taylor--couette
  flow}.  \jt{Journal of Fluid Mechanics}  \bvol{721}.

\bibitem[Osborn(1980)]{osborn1980estimates}
{\sc \au{Osborn, TR}} \yr{1980}  \at{Estimates of the local rate of vertical
  diffusion from dissipation measurements}.  \jt{Journal of physical
  oceanography}  \bvol{10}~(1),  \pg{83--89}.

\bibitem[Park {\em et~al.\/}(1994)Park, Whitehead \&
  Gnanadeskian]{park1994turbulent}
{\sc \au{Park, Young-Gyu}, \au{Whitehead, JA} \& \au{Gnanadeskian, Anand}}
  \yr{1994}  \at{Turbulent mixing in stratified fluids: layer formation and
  energetics}.  \jt{Journal of Fluid Mechanics}  \bvol{279},  \pg{279--311}.

\bibitem[Phillips(1972)]{phillips1972turbulence}
{\sc \au{Phillips, OM}} \yr{1972} Turbulence in a strongly stratified
  fluid—is it unstable?  \bt{In {\em Deep Sea Research and Oceanographic
  Abstracts\/}}, ,  \vol{vol.~19},  \pg{pp. 79--81}. Elsevier.

\bibitem[Posmentier(1977)]{posmentier1977generation}
{\sc \au{Posmentier, Eric~S}} \yr{1977}  \at{The generation of salinity
  finestructure by vertical diffusion}.  \jt{Journal of Physical Oceanography}
  \bvol{7}~(2),  \pg{298--300}.

\bibitem[Radko(2014)]{radko_2014}
{\sc \au{Radko, Timour}} \yr{2014}  \at{Applicability and failure of the
  flux-gradient laws in double-diffusive convection}.  \jt{Journal of Fluid
  Mechanics}  \bvol{750},  \pg{33–72}.

\bibitem[Radko(2016)]{radko_2016}
{\sc \au{Radko, Timour}} \yr{2016}  \at{Thermohaline layering in dynamically
  and diffusively stable shear flows}.  \jt{Journal of Fluid Mechanics}
  \bvol{805},  \pg{147–170}.

\bibitem[Radko(2019)]{radko_2019}
{\sc \au{Radko, Timour}} \yr{2019}  \at{Thermohaline layering on the
  microscale}.  \jt{Journal of Fluid Mechanics}  \bvol{862},  \pg{672–695}.

\bibitem[Rippeth \& Fine(2022)]{rippeth2022turbulent}
{\sc \au{Rippeth, Tom} \& \au{Fine, Elizabeth}} \yr{2022}  \at{Turbulent mixing
  in a changing arctic ocean}.  \jt{Oceanography} .

\bibitem[Ruddick {\em et~al.\/}(1989)Ruddick, McDougall \&
  Turner]{ruddick1989formation}
{\sc \au{Ruddick, BR}, \au{McDougall, TJ} \& \au{Turner, JS}} \yr{1989}
  \at{The formation of layers in a uniformly stirred density gradient}.
  \jt{Deep Sea Research Part A. Oceanographic Research Papers}  \bvol{36}~(4),
  \pg{597--609}.

\bibitem[Salehipour {\em et~al.\/}(2016)Salehipour, Peltier, Whalen \&
  MacKinnon]{salehipour2016new}
{\sc \au{Salehipour, H}, \au{Peltier, WR}, \au{Whalen, CB} \& \au{MacKinnon,
  JA}} \yr{2016}  \at{A new characterization of the turbulent diapycnal
  diffusivities of mass and momentum in the ocean}.  \jt{Geophysical Research
  Letters}  \bvol{43}~(7),  \pg{3370--3379}.

\bibitem[Shih {\em et~al.\/}(2005)Shih, Koseff, Ivey \&
  Ferziger]{shih_koseff_ivey_ferziger_2005}
{\sc \au{Shih, Lucinda~H.}, \au{Koseff, Jeffrey~R.}, \au{Ivey, Gregory~N.} \&
  \au{Ferziger, Joel~H.}} \yr{2005}  \at{Parameterization of turbulent fluxes
  and scales using homogeneous sheared stably stratified turbulence
  simulations}.  \jt{Journal of Fluid Mechanics}  \bvol{525},  \pg{193–214}.

\bibitem[Taylor \& Zhou(2017)]{taylor_zhou_2017}
{\sc \au{Taylor, J.~R.} \& \au{Zhou, Q.}} \yr{2017}  \at{A multi-parameter
  criterion for layer formation in a stratified shear flow using sorted
  buoyancy coordinates}.  \jt{Journal of Fluid Mechanics}  \bvol{823},
  \pg{R5}.

\bibitem[Thorpe(1982)]{thorpe_1982}
{\sc \au{Thorpe, S.~A.}} \yr{1982}  \at{On the layers produced by rapidly
  oscillating a vertical grid in a uniformly stratified fluid}.  \jt{Journal of
  Fluid Mechanics}  \bvol{124},  \pg{391–409}.

\bibitem[Timmermans {\em et~al.\/}(2008)Timmermans, Toole, Krishfield \&
  Winsor]{timmermans2008ice}
{\sc \au{Timmermans, M-L}, \au{Toole, J}, \au{Krishfield, R} \& \au{Winsor, P}}
  \yr{2008}  \at{Ice-tethered profiler observations of the double-diffusive
  staircase in the canada basin thermocline}.  \jt{Journal of Geophysical
  Research: Oceans}  \bvol{113}~(C1).

\bibitem[Turner(1968)]{turner_1968}
{\sc \au{Turner, J.~S.}} \yr{1968}  \at{The influence of molecular diffusivity
  on turbulent entrainment across a density interface}.  \jt{Journal of Fluid
  Mechanics}  \bvol{33}~(4),  \pg{639–656}.

\bibitem[Venayagamoorthy \& Stretch(2010)]{venayagamoorthy2010turbulent}
{\sc \au{Venayagamoorthy, Subhas~K} \& \au{Stretch, Derek~D}} \yr{2010}  \at{On
  the turbulent prandtl number in homogeneous stably stratified turbulence}.
  \jt{Journal of Fluid Mechanics}  \bvol{644},  \pg{359--369}.

\bibitem[Wells {\em et~al.\/}(2010)Wells, Cenedese \&
  Caulfield]{TheRelationshipbetweenFluxCoefficientandEntrainmentRatioinDensityCurrents}
{\sc \au{Wells, Mathew}, \au{Cenedese, Claudia} \& \au{Caulfield, C.~P.}}
  \yr{2010}  \at{The relationship between flux coefficient and entrainment
  ratio in density currents}.  \jt{Journal of Physical Oceanography}
  \bvol{40}~(12),  \pg{2713 -- 2727}.

\bibitem[Woods {\em et~al.\/}(2010)Woods, Caulfield, Landel \&
  Kuesters]{woods2010non}
{\sc \au{Woods, Andrew~W}, \au{Caulfield, CP}, \au{Landel, Julien~R} \&
  \au{Kuesters, A}} \yr{2010}  \at{Non-invasive turbulent mixing across a
  density interface in a turbulent taylor--couette flow}.  \jt{Journal of fluid
  mechanics}  \bvol{663},  \pg{347--357}.

\end{thebibliography}


\end{document}